\documentclass[
 reprint,
longbibliography ,
 aps,
 twoside,
 superscriptaddress,
bibnotes
]{revtex4-1}
\usepackage{amsmath, amssymb, graphicx, bm}
\usepackage{xcolor}
\usepackage{appendix}
\usepackage{algpseudocode,algorithm}
\usepackage{braket}

\newcommand{\R}{\bm{R}}
\newcommand{\A}{\bm{A}}
\newcommand{\g}{\bm{\gamma}}
\newcommand{\al}{\bm{\alpha}}
\newcommand{\sig}{\bm{\sigma}}
\newcommand{\I}{\bm{\mathrm{I}}}

\newcommand{\zero}{\bm{0}}
\newcommand{\B}{\bm{B}}
\newcommand{\bigO}{\mathcal{O}}
\newcommand{\ncut}{n_{\mathrm{cut}}}
\newcommand{\tburn}{\tau_{\mathrm{burn}}}
\newcommand{\tthin}{\tau_{\mathrm{thin}}}

\DeclareMathOperator{\lhaf}{lhaf}

\DeclareMathOperator{\Tr}{Tr}
\DeclareMathOperator{\diag}{diag}

\begin{document}

\title{
The Boundary for Quantum Advantage in Gaussian Boson Sampling
}

\author{Jacob F. F. Bulmer}
\email{these authors contributed equally:\\ jacob.bulmer@bristol.ac.uk, bryn.bell@imperial.ac.uk}
\affiliation{Quantum Engineering Technology Labs, University of Bristol, Bristol, UK}

\author{Bryn A. Bell}
\email{these authors contributed equally:\\ jacob.bulmer@bristol.ac.uk, bryn.bell@imperial.ac.uk}
\affiliation{Ultrafast Quantum Optics group, Department of Physics,
Imperial College London, London, UK}

\author{Rachel S. Chadwick}
\affiliation{Quantum Engineering Technology Labs, University of Bristol, Bristol, UK}
\affiliation{Quantum Engineering Centre for Doctoral Training, University of Bristol, Bristol, UK}

\author{Alex E. Jones}
\affiliation{Quantum Engineering Technology Labs, University of Bristol, Bristol, UK}

\author{Diana Moise}
\affiliation{Hewlett Packard Enterprise, Switzerland}

\author{Alessandro Rigazzi}
\affiliation{Hewlett Packard Enterprise, Switzerland}

\author{Jan Thorbecke}
\affiliation{Hewlett Packard Enterprise, the Netherlands}

\author{Utz-Uwe Haus}
\affiliation{HPE HPC EMEA Research Lab,  Wallisellen, Schweiz}

\author{Thomas Van Vaerenbergh}
\affiliation{Hewlett Packard Labs, HPE Belgium, Diegem, Belgium}

\author{Raj B. Patel}
\affiliation{Ultrafast Quantum Optics group, Department of Physics,
Imperial College London, London, UK}
\affiliation{Department of Physics,  University of Oxford, Oxford, UK}

\author{Ian A. Walmsley}
\affiliation{Ultrafast Quantum Optics group, Department of Physics,
Imperial College London, London, UK}

\author{Anthony Laing}
\email{anthony.laing@bristol.ac.uk}
\affiliation{Quantum Engineering Technology Labs, University of Bristol, Bristol, UK}

\begin{abstract}
Identifying the boundary beyond which quantum machines provide a computational advantage over their classical counterparts is a crucial step in charting their usefulness.
Gaussian Boson Sampling (GBS),
in which photons are measured from a highly entangled Gaussian state,
is a leading approach in pursuing quantum advantage.
State-of-the-art quantum photonics experiments that, once programmed, run in minutes, would require 600 million years to simulate using the best pre-existing classical algorithms.
Here, we present substantially faster classical GBS simulation methods, including speed and accuracy improvements to the calculation of loop hafnians, the matrix function at the heart of GBS.
We test these on a $\sim \! 100,000$ core supercomputer to emulate a range of different GBS experiments with up to 100 modes and up to 92 photons. 
This reduces the run-time of classically simulating state-of-the-art GBS experiments to several months---a nine orders of magnitude improvement over previous estimates.
Finally, we introduce a distribution that is efficient to sample from classically and that passes a variety of GBS validation methods, providing an important adversary for future experiments to test against.
\end{abstract}

\date{\today}

\maketitle

\section{Introduction}
A \emph{quantum advantage} is typically considered to be achieved when a quantum experiment outperforms a classical computer at a computational task, with strong evidence of an exponential separation between quantum and classical run-times.
Based on plausible complexity conjectures, boson sampling~\cite{aaronson2011computational, lund2014boson} is a class of photonic experiments with potential to deliver quantum advantage.
Measurement of correlated photon detection events constitutes sampling from a distribution with probabilities that correspond to classically intractable matrix functions.
In Gaussian Boson Sampling (GBS)~\cite{hamilton2017gaussian}, squeezed states are injected into an interferometer, with subsequent photon detection producing correlation events that are related to matrix \emph{loop hafnians}~\cite{PhysRevA.100.032326, quesada2019franck}.
A major advancement in experimental photonics was recently reported, in which a GBS experiment comprised of 100 optical modes, named Jiŭzhāng~\cite{Zhong1460}, observed up to 76 photon detection events
and claimed a quantum advantage. 
Once assembled, Jiŭzhāng ran in 200~s, while the best available classical algorithms running on the most powerful contemporary supercomputer would require 600~million years to simulate Jiŭzhāng.

While the theoretical proposal for GBS assumed the use of photon number resolving detectors (PNRDs), experimental implementations frequently use threshold detectors, which \emph{click} to distinguish between 0 and at least 1 photon.
This does not affect the complexity of GBS provided that \emph{collisions} (multiple photons arriving at the same detector) are unlikely~\cite{quesada2018gaussian}. Such events were assumed improbable and were neglected in the original proposal~\cite{aaronson2011computational}.
There is a lack of progress in understanding the classical complexity of GBS in regimes with high degrees of collisions,
which obfuscates the boundary for quantum advantage.
Jiŭzhāng both uses threshold detectors and operates in a regime where there is a high probability of collisions between photons.

Here, we present classical algorithms that calculate exact, correlated photon detection probabilities for GBS simulations with PNRDs, in the presence of collisions, faster than existing methods.
Futhermore, we introduce a new classical method to generate samples for GBS simulations with threshold detectors, which runs orders of magnitude faster than
classical methods to generate samples with PNRDs, when collisions dominate.
We apply these results to two sampling algorithms: a probability chain-rule method~\cite{quesada2020quadratic} and Metropolis independence sampling (MIS)~\cite{neville2017classical}. 
We report nine orders of magnitude reduction in the time taken to simulate idealised Jiŭzhāng-type GBS experiments with threshold detectors.
This enabled us to classically simulate, on a $\sim \! 100,000$ core supercomputer,
GBS experiments with 100 modes and up to 60 click detection events.
Replacing threshold detectors with PNRDs in this simulation allows us to generate a 92 photon sample but increases the run-time significantly. 
We find that simulating a 60 mode experiment with PNRDs is of comparable complexity to simulating  a 100 mode experiment with the same density of photons and threshold detectors.
Finally, we develop and investigate a classically tractable distribution that passes a variety of canonical GBS verification tests,
highlighting the importance of verifying GBS experiments against the most stringent adversarial tests available.
These results significantly sharpen the boundary of quantum advantage in GBS.

\section{Loop hafnian algorithms}
A particular detection event can be described by a photon number pattern $\vec{n}$, where $n_{i}$ is the number of photons in mode $i$. 
The probability of obtaining some $\vec{n}$ from a GBS experiment is:
\begin{equation}
    P(\vec{n})=\frac{P_0}{\prod_i n_i!}\lhaf\left(\A_{\vec{n}}\right),
    \label{mixed_prob_short}
\end{equation}
where $P_0$ is the probability of measuring vacuum, $\lhaf(\cdot)$ is the loop hafnian function, and $\A_{\vec{n}}$ is a matrix which can be derived from $\vec{n}$ and the covariance matrix and displacement vector of the Gaussian state (see Appendix~\ref{GBS}).
$\A_{\vec{n}}$ is a $2N\times 2N$ matrix, where $N=\sum_i n_i$.
However, for a pure Gaussian state, $\A_{\vec{n}}$ is block diagonal, with blocks $\B_{\vec{n}}$ and $\B_{\vec{n}}^*$, in which case:
\begin{equation}
    \lhaf\left(\A_{\vec{n}}\right)=|\lhaf(\B_{\vec{n}})|^2.
\end{equation}
$\B_{\vec{n}}$ is an $N\times N$ matrix, so it is considerably faster to calculate its loop hafnian compared to $\A_{\vec{n}}$.
While a realistic GBS experiment will not produce a pure state, a Gaussian mixed state can be expressed as a statistical ensemble of pure states with differing displacement vectors~\cite{Serafini_2017, quesada2020quadratic}; so for the purposes of a sampling algorithm, it is generally possible to randomly choose a complex displacement vector $\vec{\alpha}$ from the correct distribution, then sample from the corresponding pure state.
Hence the computational complexity of generating a sample is set by the calculation of an $N\times N$ loop hafnian, $\lhaf(\B_{\vec{n}})$.

\begin{figure}
    \centering
    \includegraphics[width=\columnwidth]{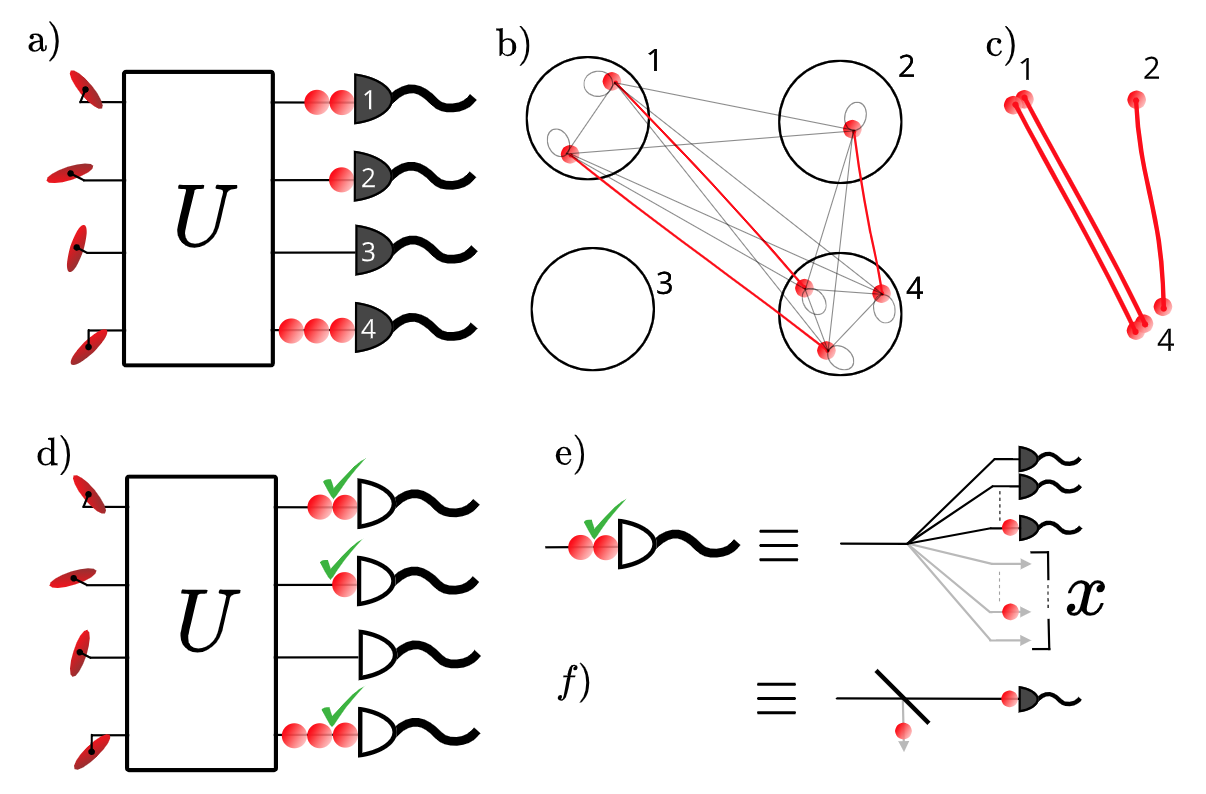}
    \caption{(a) A GBS outcome with collisions, measured with PNRDs. 
    (b) To calculate the associated probability, we group the photons into pairs (red lines) to maximise the number of repeated, identical pairs. 
    (c) An inclusion/exclusion formula, or a finite-difference sieve, can then operate on the resulting pairs, with repeated pairs leading to a speed-up. 
    (d) the same event measured with threshold detectors, with `clicks' shown as green ticks.
    (e) We consider a fan-out to an array of sub-detectors, with none likely to receive $>1$ photon.
    We can ignore the outcomes of all but the first detector to see a photon.
    $x$ is introduced as the relative position of the detected photon, and is also the fraction of the sub-detectors that are ignored.
    (f) The probability of detecting the first photon at position $x$ can be expressed as a loss followed by single photon detection.}
    \label{red_edges}
\end{figure}

The fastest known algorithms for the loop hafnian run in exponential time, using an inclusion/exclusion formula similar to the Ryser algorithm for the permanent~\cite{Ryser1963}.
In boson sampling with Fock state inputs, Ryser can be generalised to take advantage of collisions, reducing the number of inclusion/exclusion terms to calculate from $2^N$ to $\prod_i (n_i+1)$~\cite{shchesnovich2013asymptotic,tichy2011thesis, chin2018generalized}.
The repeated-moment formula for the loop hafnian achieves the same scaling for GBS~\cite{KAN2008542}.
However, there is a much faster formula for general loop hafnians---the eigenvalue-trace algorithm performs inclusion/exclusion on \emph{pairs} of photons, and so requires only $2^{N/2}$ terms~\cite{bjorklund2019faster, CYGAN201575}.
Here, we generalise eigenvalue-trace to take advantage of collisions, reducing the number of terms to $\prod_i (\eta_i+1)$, where $\eta_i$ is the number of times a particular pairing of photons is repeated.
This is lower-bounded by $\prod_i \sqrt{n_i+1}$ and upper-bounded by $2^{N/2}$.
The grouping of photons into pairs is arbitrary, so we make use of a greedy algorithm to choose repeated pairings, reducing the number of inclusion/exclusions steps to as close to the lower-bound as possible (Appendix~\ref{repeated_edges}).
Fig.~\ref{red_edges}(a)-(c) shows how, when collisions occur in more than one mode, repeated pairs can be formed.
In this example, $\vec{n}=(2,1,0,3)$, which is arranged into two pairings, one of which is repeated.
This gives a sum over $(2+1)(1+1)=6$ terms, reduced from 8 using eigenvalue-trace~\cite{bjorklund2019faster}, and compared to 24 using the repeated-moment algorithm~\cite{KAN2008542}.

We also present a loop hafnian formula which uses a \emph{finite difference sieve} instead of an inclusion/exclusion formula, like the Glynn formula for the permanent~\cite{Balasubramanian1980thesis, BAX1996171, GLYNN20101887}.
This significantly improves the numerical accuracy with only a minor time penalty.
Whereas accuracy is an issue for eigenvalue-trace when $N>50$~\cite{quesada2020quadratic}, the finite-difference sieve method has relative error $<10^{-8}$ when tested up to $N=60$ (Appendix~\ref{fds}).
This allows us to maintain accuracy for large loop hafnians while using a conventional 128-bit complex floating point data type, which is desirable for speed and portability.
We therefore use the finite difference sieve formula for all benchmarking results presented in section~\ref{benchmarking}. 

\section{Threshold detectors}
When threshold detectors are used, the detection probabilities can be calculated using the Torontonian matrix function~\cite{quesada2018gaussian}, which involves a sum over $2^{N_c}$ terms, where $N_c$ is the number of clicks (outputs with one or more photons).
However, calculating this quantity is not necessarily the fastest approach to \emph{sampling} threshold detection patterns.
For a sufficiently low density of photons, it may be faster to simulate PNRDs, then simply reduce each non-zero photon number to a click.
We show that it is possible to improve this, for any density of photons, to the level of an $N_c\times N_c$ loop hafnian, containing $2^{N_c/2}$ terms.

We consider the detection system depicted in Fig.~\ref{red_edges}(e).
The mode is uniformly fanned out to many PNRD sub-detectors, such that the probability of a collision in any one sub-detector can be neglected.
This system provides a conceptual bridge between threshold detection and number resolved detection~\cite{thomas2020general}.
If these sub-detectors within a mode are sampled sequentially, once a single photon is seen, that mode registers a click.
The remaining sub-detectors, which have not yet been sampled, can be ignored since no more information is required about that mode.
Hence the number of detected single photons to simulate is $N_c$, which sets the size of loop hafnian calculation.
$x$ is introduced as an additional variable giving the position of the single photon within the fan-out, normalised to vary between 0 and 1.
As a result, a fraction $x$ of the sub-detectors are ignored - this can be related to applying a loss of $x$ to the mode before detecting a single photon, shown in Fig.~\ref{red_edges}(f).

\section{Sampling algorithms}
\subsection{Chain-rule sampling}
These methods can be applied directly to the chain-rule for simulating GBS described in~\cite{quesada2020quadratic} and Appendix~\ref{chain_rule}.
Here, the photon number in each mode is sampled sequentially, conditioned on the photon numbers in the previous modes.
Finding the conditional probability distribution for mode $j$ requires calculating the joint probabilities of $(n_1, ..., n_j)$ for all values of $n_j$ up to $\ncut$, where $\ncut$ is some cutoff such that the probability of having a greater number of photons can be neglected.
Since $\ncut$ should generally be several times larger than the expected number of photons, the speed-up for calculating collision probabilities is especially applicable here.
Furthermore, we make use of a batched method for simultaneously calculating all of the loop hafnians required for different values of $n_j$, with approximately the same run-time as calculating the largest loop hafnian, where $n_j=\ncut$ (see Appendix~\ref{batching}).
When simulating threshold detectors, we choose to reduce $\ncut$ for each sub-detector to 1.
We again use a batching method to more efficiently sample different sub-detectors within the same mode, which largely offsets the additional overhead from sampling several sub-detectors per mode.

\subsection{Metropolis Independence Sampling}
We also investigate MIS, a Markov Chain Monte Carlo method, for generating GBS samples.
Here, samples $s_i$ are drawn from a proposal distribution, where $s_i$ is the $i$th sample in the chain. They are then accepted with probability
\begin{equation}
    p_{\mathrm{accept}}=\mathrm{min}\left(1,\frac{P(s_i)Q(s_{i-1})}{P(s_{i-1})Q(s_i)}\right),
    \label{eq:transitionprob}
\end{equation}
where $P(s_{i})$ is the \emph{target} probability distribution, in this case that of ideal GBS, while $Q(s_{i})$ is the proposal probability distribution, i.e. the probability of proposing a particular $s_i$.
If a proposed sample is rejected, the previous sample is repeated, $s_i=s_{i-1}$.
This update rule ensures the chain will converge towards the target distribution, which is its equilibrium state~\cite{liu1996metropolizedindependent, 10.5555/1571802}.
Usually some burn-in time, $\tburn$, is used to allow the chain to converge.
As sequential samples are not independent, some thinning interval, $\tthin$, can also be used to suppress the probability of seeing repeated samples, keeping only 1 in every $\tthin$ samples.
These parameters are critical to the efficiency of MIS, and can generally be improved by choosing a proposal distribution which is close to the target distribution.

\begin{figure}[t]
    \centering
    \includegraphics[width=\columnwidth]{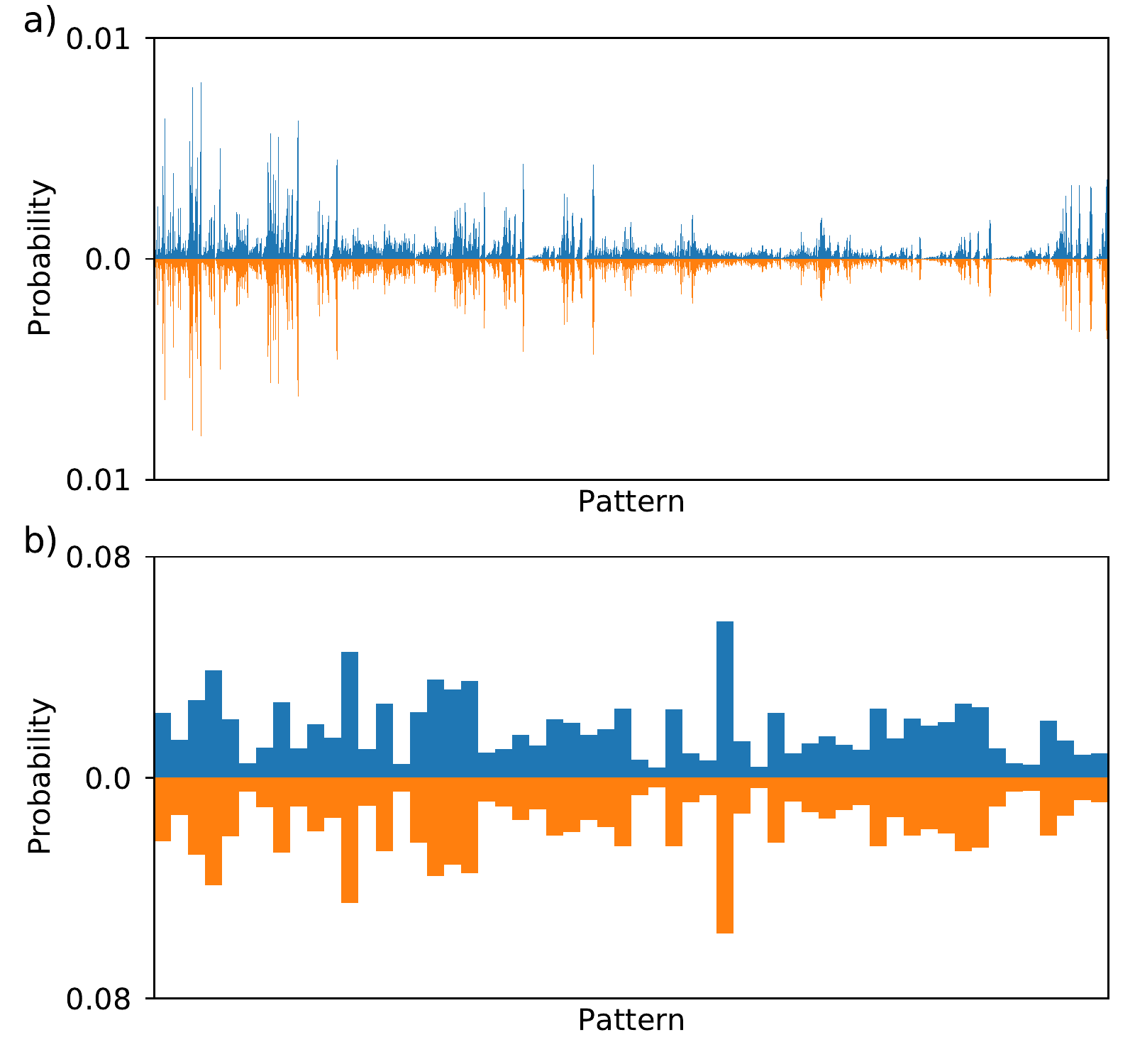}
    \caption{Probability distribution for all 6-photon detection outcomes for an 8-mode PNRD GBS simulation (a) and an 8-mode, 3-click threshold detector GBS simulation (b). Blue bars show estimated probabilities using MIS, orange bars show exact probabilities.}
    \label{pnrd_dist}
\end{figure}

We expand our sample space so that $s_i$ contains the photon number pattern $\vec{n}$ and the complex displacement vector $\vec{\alpha}$.
For the proposal samples, we draw $\vec{\alpha}$ from the correct distribution for the desired mixed state, then generate $\vec{n}$ from an `Independent Pairs and Singles' (IPS) distribution based on the resulting pure state (Appendix~\ref{proposal}).
This distribution, that we introduce in this work, can be sampled from efficiently, and has probabilities given by $N\times N$ loop hafnians of positive matrices.
As an aside, we observe that the IPS distribution is already sufficient to pass many GBS verification methods (Appendix~\ref{verification}). 
The run-time per sample is dominated by the two loop hafnians in $P(s_i)$ and $Q(s_i)$, with $P(s_{i-1})$ and $Q(s_{i-1})$ already calculated in the previous step.

When simulating threshold detectors with MIS, we take the continuum limit of a large number of sub-detectors, and introduce $x$ as an additional continuous random variable that gives the position of the `first' detected photon within each mode with non-zero photons.
Given a proposed photon number pattern, $\vec{x}$ can be sampled efficiently from its conditional distribution $p(\vec{x}|\vec{n})$.
$P(s_i)$, $Q(s_i)$ are then calculated with $N_c\times N_c$ loop hafnians, tracing out the unused sub-detectors.
One subtlety is that tracing out reintroduces mixture into the quantum state, so it is necessary to sample a further adjustment to the displacement vector $d\vec{\alpha}$ to obtain a pure state.
This is only used in the calculation of $P(s_i)$. Details are given in Appendix~\ref{click_mis}.

\begin{figure}[t]
    \centering
    \includegraphics[width=\columnwidth]{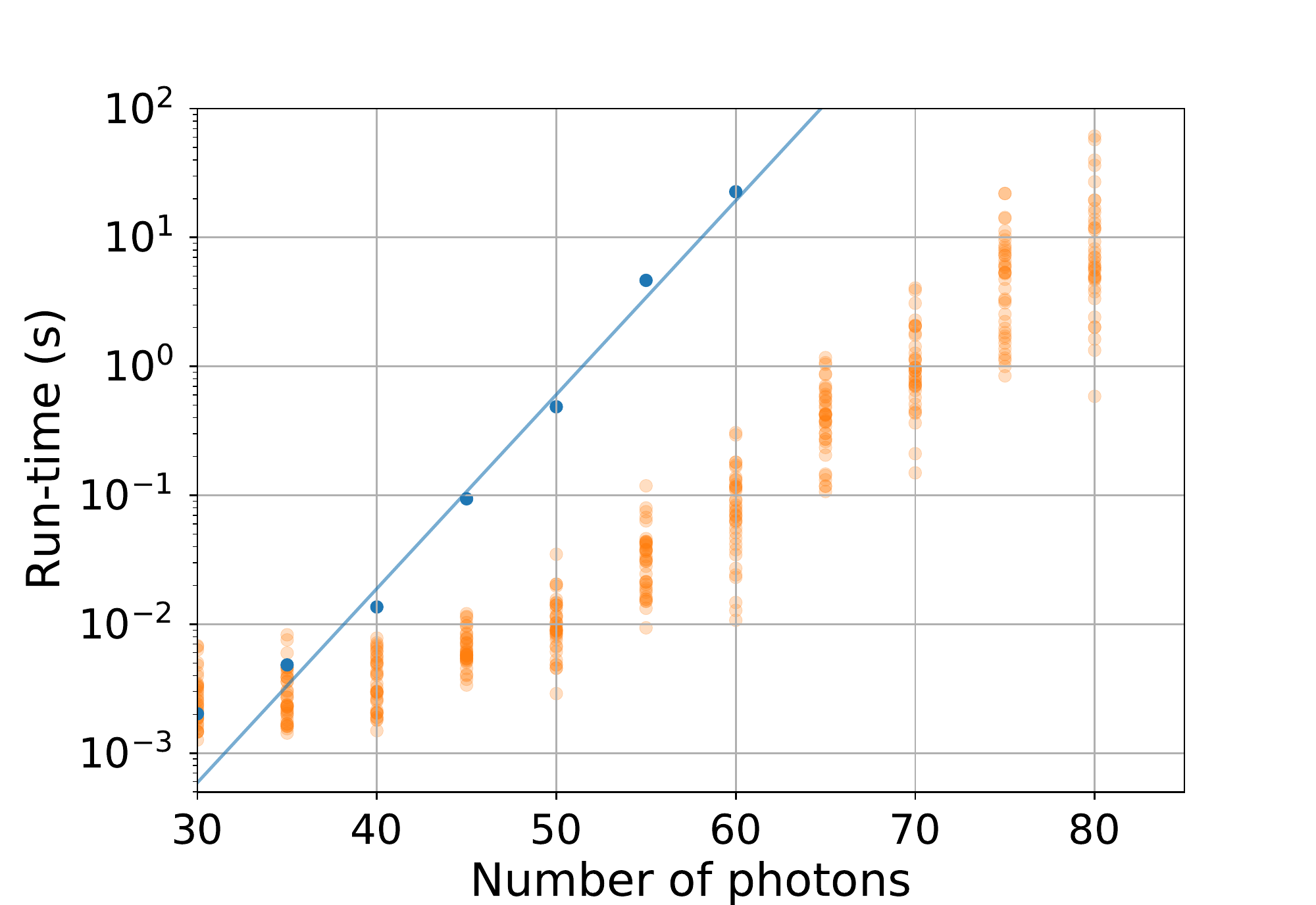}
    \caption{Run-time using the HPE benchmarking system, comparing eigenvalue-trace loop hafnian algorithm on $N \times N$ matrices with and without speed-up due to collisions (orange and blue dots). Blue line is an exponential fitted to the blue points. Collisions are determined by generating 39 samples for each $N$ from the IPS distribution on 60 modes.}
    \label{speed_repeats}
\end{figure}

\begin{figure}[ht]
    \centering
    \includegraphics[trim={0cm 1.5cm 1.4cm 2.5cm},clip,width=\columnwidth]{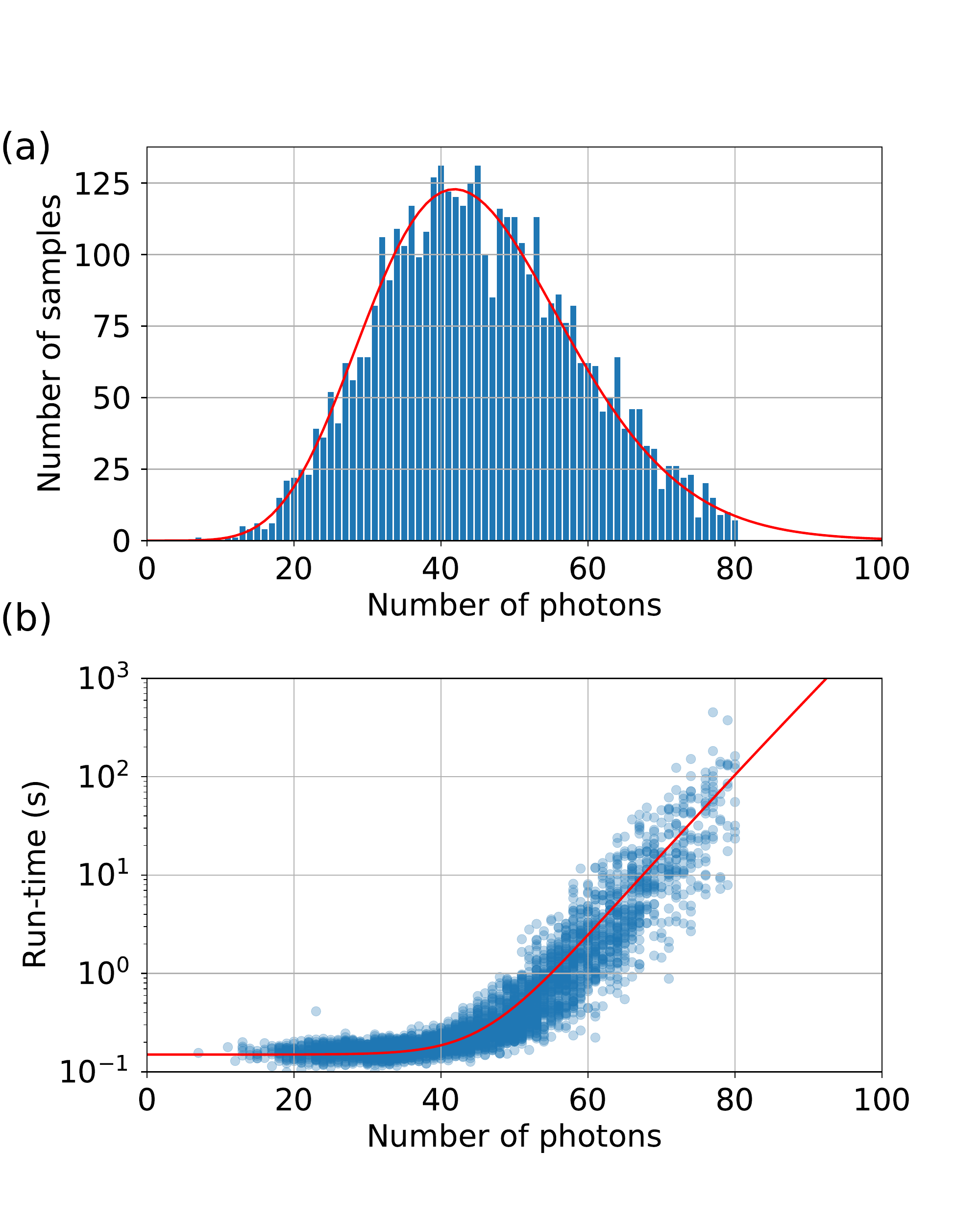}
    \caption{Chain-rule simulation of $M=60$ experiment with PNRDs. (a) Number of samples as a function of photon number, with the theoretically calculated distribution (red line) and (b) run-time versus number of photons fitted with an exponential plus a constant (red line).}
    \label{chainrulePNRD}
\end{figure}

\section{Benchmarking \label{benchmarking}}
To benchmark these methods, we choose parameters similar to those of Zhong et al.~\cite{Zhong1460}, while varying the system size by choosing the number of modes $M$.
For the interferometer we select Haar random unitary matrices, fed with $M/4$ sources of two-mode squeezed vacuum.
We choose a uniform squeezing parameter $r=1.55$, and overall transmission $\eta=0.3$.
To demonstrate the correctness of our methods, we first test them on an $M=8$ example, which is small enough that the results can be compared to the exactly calculated distributions.
Fig.~\ref{pnrd_dist} shows the accumulated distribution from $10^6$ samples with total photon number $N=6$,
generated by MIS for PNRD, along with the exactly calculated distribution.
The total variation distance,
$\mathrm{TVD}(p,q) = \frac12 \sum_i |p_i-q_i|=0.0153$, which is consistent with statistical uncertainties.
With threshold detectors, we find the TVD for the $N_c=3$ distribution is $2.9\times 10^{-3}$, which benefits from the smaller statistical uncertainty due to the smaller number of possible outcomes.
For the chain-rule algorithm, we produce $10^6$ samples with both PNRD and threshold detectors. 
For PNRD with a cutoff of 12 photons, there were 74,973 samples with $N=6$ from $10^6$ total samples, giving a $\mathrm{TVD}=0.0554$. 
For threshold detectors with twelve sub-detectors, there were 195,150 $N_c=3$ samples and these gave a $\mathrm{TVD}=0.0138$.
The larger TVDs are explained by the smaller sample size of the post-selected distributions.

For large-scale tests we make use of an internal HPE Cray EX benchmarking system, consisting of 1024 nodes.
A typical node is equipped with two AMD EPYC 7742 64-core processors clocked at 2.25GHz and the nodes are interconnected with the Cray Slingshot 10 high-performance network.
We first benchmark our loop hafnian formula on proposed IPS samples for an $M=60$ example. 
The run-time as a function of $N$ is shown in Fig.~\ref{speed_repeats}, along with timings for the basic formula without speed-up due to collisions.
Making use of collisions generally improves the run-time by one to two orders of magnitude for this range, and allows 80 photon probabilities to be calculated in comparable time to a 60 photon probability without collisions.
However, there is a large variation in run-time between samples with the same $N$, depending on the amount of collisions in any particular configuration of the sampled photons. 
On the other hand, the run-time for a loop hafnian without speed-up from collisions shows little variation from $\mathcal{O}(N^3 2^{N/2})$ scaling, at least for $N>40$.

\begin{figure}[t]
    \centering
    \includegraphics[trim={0cm 1.5cm 1.5cm 2.5cm},clip,width=\columnwidth]{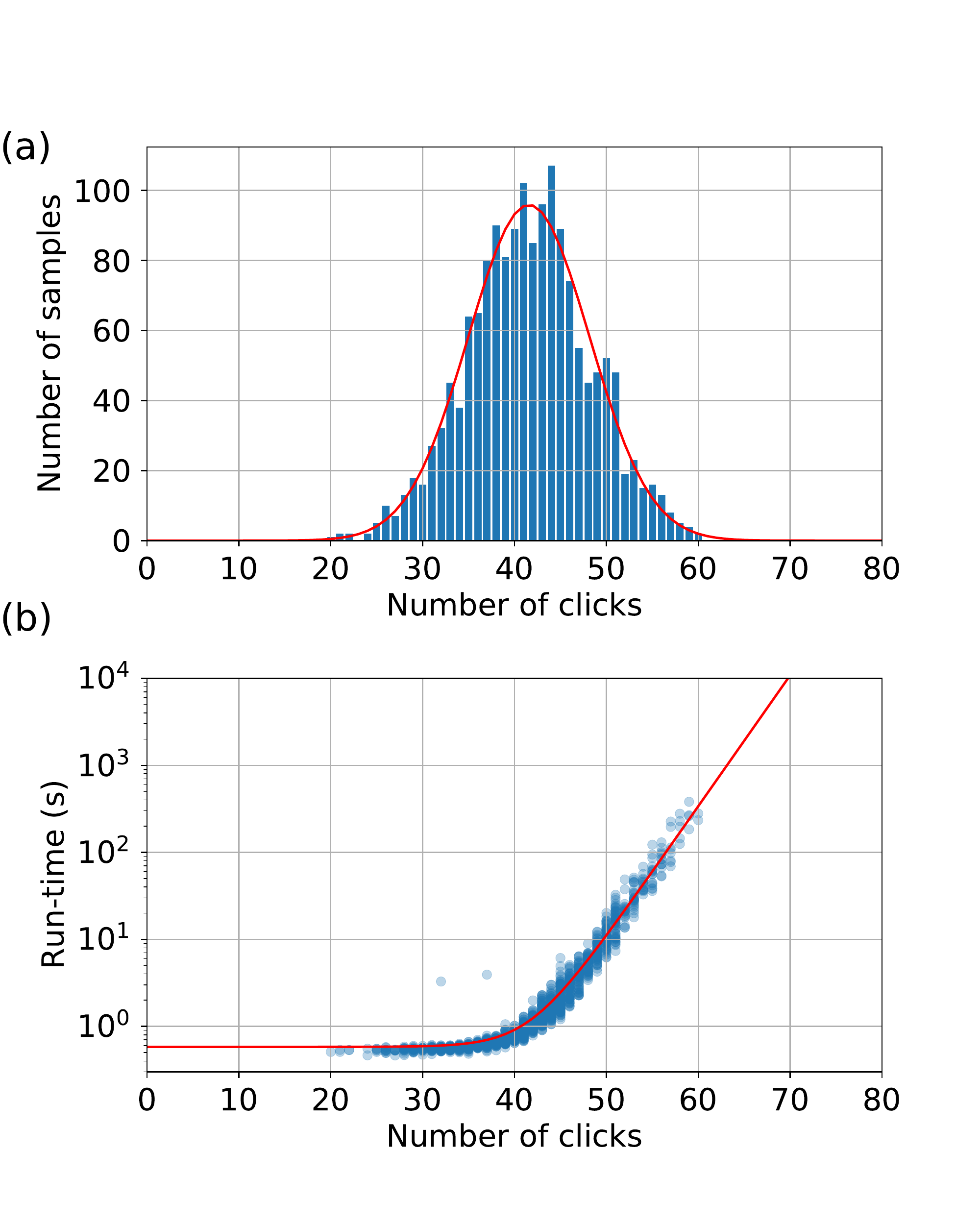}
    \caption{Chain-rule simulation of $M=100$ experiment with threshold detectors. (a) Number of samples as a function of number of clicks, fitted with a Gaussian (red line) (b) Run-time versus number of clicks, fitted with an exponential plus a constant (red line).}
    \label{chainruleclick}
\end{figure}

Using chain-rule sampling, we simulate an $M=60$ experiment with PNRDs, setting $\ncut=12$ and an additional global cutoff of $80$ photons.
We generate 4200 samples in $\sim \! 3$ hours.
The global cutoff has no effect on the probability distribution of samples below the cutoff, and is used to keep the run-time per sample constrained.
Fig.~\ref{chainrulePNRD}(a) shows a histogram of the number of samples against number photons, which is in good agreement with the calculated distribution.
Fig.~\ref{chainrulePNRD}(b) shows the corresponding run-times of the samples.
Below $\sim \! 45$ photons, the sample time appears approximately constant, suggesting the problem size is not large enough to take full advantage of the system.
Beyond that, the run-time increases rapidly, though there is a wide range of variation depending on the particular configuration of output photons.
We provide a rough fit-line to this scaling, equal to $(0.15+1.59\times 10^{-9}\times N^3e^{0.147 N})$s. 
Using this to extrapolate to photon numbers $>80$, we estimate that the average time per sample is $\sim \! 10$~s.
With the $\sim \! 66$ times larger number of CPUs available in Fugaku---the world's top ranked supercomputer---this could be reduced to 130~ms. 

We then test chain-rule simulation of an $M=100$ experiment with threshold detectors, using 12 sub-detectors per mode, and a global cutoff of $60$ clicks.
We generate 1600 samples in $\sim \! 3.5$ hours.
Fig.~\ref{chainruleclick}(a) shows the histogram of click numbers, and (b) shows the corresponding run-times of the samples.
Beyond $\sim \! 45$ clicks, the sample time increases approximately exponentially, from which we extrapolate to click numbers $>60$.
The run-times are fitted with a line $(0.58+3.15\times 10^{-7}\times 2^{N/2})$s.
From this, we predict that the mean time per sample is 8.4~s.
On Fugaku this could be reduced to around 127~ms.

Based on the scaling of the loop hafnian calculation, and on the distribution of samples over number of clicks, the estimated average time per MIS step is 0.45~s for an $M=100$ system with threshold detectors.
On Fugaku, this could be reduced to 7~ms, which is somewhat faster than generating a sample through the chain-rule.
However, the raw MIS chain will contain a high frequency of repeated samples due to rejections of the proposal sample---for some applications this may be unimportant, but it would provide a clear difference from a true GBS experiment, where repeated samples are highly unlikely.
In Appendix~\ref{mis_scaling} we investigate the $\tthin$ required to suppress repeated samples, and find it increases rapidly with system size, such that for $M=100$ it is likely to be in excess of $600$.
Hence if independently distributed samples are required, the chain-rule method is most likely preferable.

\section{Conclusion}

Our results provide a new reference point for classical run-times of GBS,
an improved understanding of the classical complexity,
and could improve verification techniques by making it practical to generate small numbers of samples from the distribution of much larger scale experiments.
Our `Independent Pairs and Singles' proposal distribution generates samples in polynomial time and is a better approximation than the standard adversarial models in the verification of GBS.
IPS is largely able to pass the quantitative tests of GBS used in ref.~\cite{Zhong1460} (see Appendix~\ref{verification}), which suggests a need for stronger verification methods - at the least, using IPS as a classically simulable adversary.

For GBS with threshold detectors, we have shown the complexity can be reduced quadratically from $\mathcal{O}(N_c^3 2^{N_c})$ to $\mathcal{O}(N_c^3 2^{N_c/2})$.
Comparing to the experiment of Zhong et al.~\cite{Zhong1460}---where 50 million samples were accumulated in 200~s---our 100 mode chain-rule simulation implies the classical run-time can be reduced to $\sim \! 73$ days.
This does not diminish the experimental achievement of large-scale GBS from Zhong et al.~\cite{Zhong1460}, which remains faster than classical methods on supercomputers, if the time required for circuit programming (or in the present case fabricating a new, fixed interferometer) is not included.
However, it has previously been reported that in boson sampling with Fock state inputs, at least 50 photon events are required to extend beyond the reach of an exact classical simulation in a reasonable time-scale~\cite{neville2017classical}; for collision-free GBS, this threshold has been reported as being around 100 photons~\cite{quesada2020quadratic}; we have now demonstrated that for GBS with threshold detectors, the number of correlated detector clicks should also be around 100. 
For GBS with PNRDs, the number of photons required to surpass this classical threshold will depend on the amount of collisions, but must be $\ge 100$.

Future claims to quantum advantage in GBS experiments might include increasing the level of programmability~\cite{zhong2021phaseprogrammable} and including photon number resolving detectors, which our results suggest adds significantly to the complexity, thus providing an alternative route to a larger quantum advantage than increasing the size of threshold detector experiments.
For example, our 60 mode PNRD chain-rule simulation ran in comparable time to the 100 mode threshold detector simulation.
Meanwhile a 100 mode PNRD simulation proved impractically slow even on the HPE Cray EX benchmarking system - we generated a single 92 photon event in 82 minutes.
Our methods are near-exact simulations of GBS which do not assume or exploit any experimental imperfections beyond the presence of collisions, and so are quite generally applicable to future GBS experiments. Much faster classical methods to simulate GBS may be possible through other techniques that exploit errors such as photon loss and photon distinguishability~\cite{renema2018,garciapatron2019}, or limitations such as the inability to implement Haar random transformations.

\section*{Acknowledgements}
We thank Stefano Paesani, Alex Neville, Oliver Thomas, Dara McCutcheon, Nicol\'{a}s Quesada and Ray Beausoleil for helpful discussions. 
We acknowledge The Walrus python library as a helpful reference on loop hafnian algorithms~\cite{Gupt2019}. 
This work used the Isambard 2 UK National Tier-2 HPC Service (http://gw4.ac.uk/isambard/) operated by GW4 and the UK Met Office, and funded by the Engineering and Physical Sciences Research Council (EPSRC) (EP/T022078/1).
We acknowledge support from the EPSRC Hub in Quantum Computing and Simulation (EP/T001062/1) and Networked Quantum Information Technologies (EP/N509711/1).
R.S.C acknowledges support from EPSRC (EP/LO15730/1). 
Fellowship  support  from  EPSRC  is acknowledged by A.L. (EP/N003470/1).
B.A.B. is supported by a European Commission Marie Skłodowska Curie Individual Fellowship (FrEQuMP, 846073).

\appendix 
\section{Gaussian Boson Sampling}\label{GBS}

The Wigner function of an $M$-mode Gaussian state can be efficiently represented by using the $2M$ length mean vector $\R$, and the $2M\times2M$ covariance matrix $\bm{V}$, of the canonical position and momentum operators $\vec{q}$ and $\vec{p}$. 
Equivalently it can be represented in terms of creation and annihilation operators $\bm{a}=\left( \begin{smallmatrix} \vec{a} \\ \vec{a}^\dagger \end{smallmatrix}\right)$ as a complex valued displacement $\al$ and covariance matrix $\sig$~\cite{RevModPhys.84.621}.
\begin{equation}
    \al_i=\braket{\bm{a}_i}
\end{equation}
\begin{equation}
    \sig_{i,j}=\frac{1}{2}\left(\braket{\bm{a}_i \bm{a}_j^\dagger} + \braket{\bm{a}_j^\dagger \bm{a}_i}\right)-\al_i\al_j^*.
\end{equation}
We further define: $\sig_Q=\sig + \I/2$ as the complex-valued covariance matrix of the state's Husimi Q-function, $\bm{O}=\left(\I - \sig_Q^{-1}\right)$,
\begin{equation}
    \bm{X}=\begin{pmatrix}\zero & \I \\ \I & \zero\end{pmatrix},
\end{equation}
$\bm{A}=\bm{X}\bm{O}$, and $\g = \al^\dagger \sig_Q^{-1} $.

Probabilities of measuring photon number patterns $\vec{n}$ with PNRDs are now given by:
\begin{equation}
    P(\vec{n}|\sig,\al)=\frac{\exp\left(-\frac{1}{2} \al^\dagger \sig_Q^{-1} \al \right)}{\sqrt{\det(\sig_Q)}\prod_i n_i!}\lhaf\left(\A_{\vec{n}}\right),
    \label{mixed_prob}
\end{equation}
where $\lhaf(\cdot)$ is the loop hafnian function. $\A_{\vec{n}}$ is formed from $\A$ by repeating the $i$th and $(i+M)$th rows and columns $n_i$ times, and similarly the $i$th and $(i+M)$th entry in $\g$ is repeated $n_i$ times to form $\g_{\vec{n}}$. Then the diagonal elements of $\A_{\vec{n}}$ are replaced by the elements of $\g_{\vec{n}}$, since the weights of the loops are given on the diagonal of the matrix.

For pure states, $\A_{\vec{n}}$ can be written in block form as
\begin{equation}
    \A_{\vec{n}} = \begin{pmatrix}
    \B_{\vec{n}} & \zero \\
    \zero & \B_{\vec{n}}^*
    \end{pmatrix}.
    \label{Adef}
\end{equation}
Here $\B_{\vec{n}}$ is a symmetric $N\times N$ matrix, with $N$ the total photon number. As a result,
\begin{equation}
    \lhaf\left(\A_{\vec{n}}\right)=\left|\lhaf\left(\B_{\vec{n}}\right)\right|^2,
\end{equation}
so probabilities from a pure state can be calculated using loop hafnians of matrices of half the size compared to a mixed state.

\subsection{Sampling pure Gaussian states from mixed Gaussian states \label{sample_pure}}

Using the Williamson decomposition, we can write the covariance matrix as, $\bm{V} = \bm{S} \bm{D} \bm{S}^T$.
Here, $\bm{D}$ is a diagonal covariance matrix describing a thermal state in each mode, and $\bm{S}$ defines a symplectic transformation. 
Hence any mixed Gaussian state can be written as a pure channel acting on thermal states. 

By defining $\bm{T}=\frac{\hbar}{2}\bm{S}\bm{S}^T$, a covariance matrix of a pure Gaussian state, and $\bm{W} = \bm{S}(\bm{D} - \frac{\hbar}{2}\I)\bm{S}^T$, a covariance matrix describing the Gaussian classical noise added to the state, we can now write the original covariance matrix as $\bm{V}=\bm{T} + \bm{W}$~\cite{Serafini_2017, quesada2020quadratic}.

For the purposes of sampling the state, we can choose a pure state with vector of means $\bm{R'}$ sampled from the multivariate normal distribution described by covariance matrix $\bm{W}$ and means $\bm{R}$.
This results in a pure state with covariance matrix given by $\bm{T}$ and means given by $\bm{R'}$.

\subsection{Chain-rule GBS sampler\label{chain_rule}}

Sampling using the chain-rule for probability proceeds by choosing part of the sample (in this case, e.g. the number of photons in the first mode) from its marginal probability distribution, then fixing this and choosing the next part (e.g. number of photons in the second mode) from its conditional probability distribution depending on the first part.
This is expressed as:
\begin{equation}
    P(n_1,n_2)=P(n_1)P(n_2|n_1).
\end{equation}
This allows samples to be built-up from distributions with very large numbers of possible outcomes, without calculating the probability of every possible outcome.
In GBS, a difficulty is that the marginal probabilities are equivalent to probabilities from a mixed quantum state, and these are quadratically harder to calculate than for a pure state.
To circumvent this, the modes are initially sampled in the coherent state basis, obtaining a set of coherent state amplitudes $\vec{\beta}$ which are then progressively replaced by photon numbers $\vec{n}$ using a modified form of the chain-rule.
The coherent state basis has the benefits that it can be sampled from efficiently, and that when intermediate probabilities are calculated, combining photon number and coherent state bases, the coherent states do not add to the complexity of the calculation.
The procedure is as follows~\cite{quesada2020quadratic}:
\begin{enumerate}
    \item Sample modes 2 to $M$ in the coherent state basis, obtaining a sample from the distribution $P(\beta_2,...,\beta_M)$.
    \item Sample the photon number in the first mode from the distribution $P(n_1|\beta_2,...,\beta_M)$.
    \item For $m=2$ to $M-1$:
    \begin{enumerate}
        \item Begin with a sample from the intermediate distribution. $P(n_1,...,n_{m-1},\beta_m,...,\beta_M)$
        \item Discard the coherent state amplitude $\beta_m$ and replace it with a photon number $n_m$ drawn from the distribution $P(n_j|n_1,...,n_{m-1},\beta_{m+1},...,\beta_M)$.
        \item This leaves a sample drawn from the distribution $P(n_1,...,n_m,\beta_{m+1},...,\beta_M)$ which can be used as a starting point for the next step.
    \end{enumerate}
    \item Discard $\beta_M$ and replace it with $n_M$, drawn from $P(n_M|n_1,..,n_{M-1})$. This leaves a photon number sample drawn from the distribution $P(n_1,...,n_M)$.
\end{enumerate}

To sample from $P(n_m|n_1,...,n_{m-1},\beta_{m+1},...,\beta_M)$, the joint probabilities $P(n_1,...,n_m,\beta_{m+1},...,\beta_M)$ are calculated for all $n_m$ between zero and some finite cutoff $\ncut$.
Assuming the probability that $n_j>\ncut$ is small enough to be neglected, normalising the joint probabilities to 1 provides a good approximation to the conditional distribution.
Calculating these joint probabilities dominates the computational effort for sampling each mode, and grows with the number of detected photons.
Specifically, the relative joint probabilities are given by:
\begin{equation}
    P(n_1,...,n_m,\beta_{m+1},...,\beta_M)\propto \frac{\lhaf(\B_{\vec{n},\vec{\beta}})}{n_m!},
\end{equation}
where $\B_{\vec{n},\vec{\beta}}$ is formed from $\B$ by repeating the $i$th row and column $n_i$ times, then in the same manner repeating the entries of $\gamma'$ along the diagonal of $\B_{\vec{n},\vec{\beta}}$, where $\gamma'$ is given by:
\begin{equation}
    \gamma'=(\vec{\alpha}-\vec{\beta})^\dagger \bm{\sigma}_Q^{-1}.
\end{equation}
Here, $\vec{n}$ is non-zero only for the modes which have already been sampled in photon number, and similarly the values of $\vec{\beta}$ are set to zero as the corresponding mode is sampled in photon number.
We note that since $\ncut$ should usually be several times greater than the expected number of photons, these calculations will often contain photon collisions.
Below, we describe algorithms to speed up loop hafnian calculations in the presence of detecting photon collision events, and a method of batching together the calculations for different $n_j$ such that the total run-time is approximately equal to that of calculating the largest $n_j$.

When simulating threshold detectors, we expand each mode to several sub-detectors and treat them as separate modes in the chain-rule sampling algorithm, with the only difference being that once a photon is detected, no further information is required from the remaining sub-detectors within that mode.
Hence they can continue to be projected onto the coherent state basis, where they do not contribute to the complexity of calculating the probabilities.
In section~\ref{batching}, we provide a batched method of calculating the loop hafnians required for different sub-detectors within the same mode, achieving a speed-up by noting that only the diagonal entries of $\B_{\vec{n},\vec{\beta}}$ change between sub-detectors.

Since the order with which this algorithm progresses through the modes is arbitrary, we choose to go in order of increasing mean photon/click number. 
This slightly reduces the run-time since photons are less likely to be detected in the earlier modes, and so the size of the loop hafnians required in these stages is generally reduced.
An implementation of the chain-rule algorithm can be found in~\cite{gbs_mis}.

\section{Loop hafnian algorithms}
The loop hafnian function of an $N \times N$ symmetric matrix $A$ is defined as
\begin{equation}
    \lhaf(A)=\sum_{M\in \mathrm{SPM}}\prod_{(i,j)\in M} A_{i,j},
\end{equation}
where $\mathrm{SPM}$ is the set of single-pair matchings, the ways in which the indices $[N]$ can be grouped into sets of sizes 1 and 2. This is a generalisation of the set of perfect matchings (all of the groupings into pairs) which occurs in a hafnian, with the `loops' referring to sets of size 1, which have weightings given on the diagonal of the matrix. Hence $M$ can contain pairs $(i,j)$ where $i\neq j$, but also $(i,i)$ singles.

\subsection{Eigenvalue-trace}

The eigenvalue-trace algorithm for the loop hafnian (with $N$ even) can be written as~\cite{bjorklund2019faster}:
\begin{equation}
    \lhaf(A) = \sum_{Z \in P([N/2])} (-1)^{|Z|} f\left(A_Z\right).
    \label{ET_loop}
\end{equation}
$P([N/2])$ is the powerset of $[N/2]$, and subscript $Z$ refers to taking a submatrix where rows and columns $i$ and $N/2+i$ are retained only if $i$ is an element of the set $Z$.
The function $f(C)$ is defined as the $\lambda^{N/2}$ coefficient of the polynomial:
\begin{multline}
    p_{N/2}(\lambda, C,v) = \\
    \sum_{j=1}^{N/2} \frac{1}{j!} \left(\sum_{k=1}^{N/2} \left( \frac{\Tr((CX)^k)}{2k} + \frac{v X(CX)^{k-1} v^T}{2} \right) \lambda^k \right)^j
    \label{ET_poly}
\end{multline}
where $v$ is a vector given by the diagonal elements of $C$ and $X$ is defined like $\bm{X}$, introduced earlier, but with dimensions matching $C$.
The eigenvalue-trace algorithm can be thought of as performing inclusion/exclusion over the set of pairs in one fixed perfect matching, defined by $X$.

The complexity of evaluating $f(C)$ is dominated by finding the traces of matrix powers, $\Tr((CX)^k)$, which can be reduced to finding the eigenvalues of $CX$ in $\mathcal{O}(N^3)$ time.
Given there are $2^{N/2}$ terms in the summation in Eq.~\ref{ET_loop}, this results in $\bigO(N^3 2^{N/2})$ complexity.

\subsection{Repeated pairs \label{repeated_edges}}

This algorithm makes use of a fixed perfect matching given by the adjacency matrix $X$, defining pairs $(i,N/2+i)$ for $i\in [1,N/2)$.
The summation in Eq.~\ref{ET_loop} corresponds to inclusion/exclusion of these pairs.
If we consider the way that the $\A_{\vec{n}}$ matrix is formed when evaluating the $\vec{n}$ probability from a mixed state, $X$ will pair the $i$th index in $\A$ with the $(i+M)$th index, and this pairing will be repeated $n_i$ times.
Instead of summing over all inclusion/exclusion possibilities, we can sum over a vector $\vec{z}$ where $z_i$ runs from 0 to $n_i$, corresponding to including $z_i$ copies of the $i$th pair:
\begin{equation}
    \mathrm{lhafmix}(\A,\g,\vec{n})=\sum_{\vec{z}} (-1)^{|\vec{z}|}\prod {n_i \choose z_i} f'(\A, \g, \vec{z}).
\end{equation}
We label this function $\mathrm{lhafmix}$ because it does not apply to general matrices, only those with the particular form of $\A_{\vec{n}}$.
$f'(C, \vec{v}, \vec{z})$ is defined as the $\lambda^{N/2}$ coefficient in the polynomial
\begin{multline}
    p'_{N/2}(\lambda, C,\vec{v},\vec{z}) = \\
    \sum_{j=1}^{N/2} \frac{1}{j!} \left(\sum_{k=1}^{N/2} \left( \frac{\Tr((CX_{\vec{z}})^k)}{2k} + \frac{v X_{\vec{z}}(CX_{\vec{z}})^{k-1} v^T}{2} \right) \lambda^k \right)^j
\end{multline}
where
\begin{equation}
    X_{\vec{z}} = \begin{pmatrix}
    \zero & \diag(\vec{z}) \\
    \diag(\vec{z}) & \zero
    \end{pmatrix},
\end{equation}
with $\diag(\vec{z})$ a diagonal matrix containing the elements of $\vec{z}$.
This makes use of the fact that increasing the weight of a pairing in $X$ has the same effect as including a pair multiple times.
Where there are elements $z_i=0$, the $i$th and $(N/2+i)$th row/column can be deleted from $\A$ and $X$ to speed up the eigenvalue calculation.

This algorithm calculates mixed state probabilities in time $\mathcal{O}\left(N^3\prod_i (n_i+1)\right)$.
Noting that this corresponds to a $2N\times 2N$ loop hafnian, this compares well to using the repeated moment algorithm~\cite{KAN2008542}, which would take $\mathcal{O}\left(N\prod_i (n_i+1)^2\right)$, and improves on eigenvalue-trace whenever there are elements of $\vec{n}$ greater than 1.

For general matrices such as those in pure state calculations, even if there are photon collisions which lead to repeated rows/columns in the $B$ matrix, these do not necessarily lead to repeated pairings.
However if identical pairs do occur, with the $i$th pair occurring $\eta_i$ times, we can make use of the above formula to obtain some speed-up, reducing the number of inclusion/exclusion terms to $\prod_i (\eta_i+1)$.
This quantity is upper-bounded by $2^{N/2}$, which occurs if no pairs are repeated.
It is lower-bounded by $\prod_j \sqrt{n_j+1}$.
To see this, consider that for a total of $H$ unique pairings we can write:
\begin{equation}
    \prod_{i=1}^H (\eta_i+1) = \prod_{i=1}^H \prod_{j=1}^M \prod_{k=L,R} \sqrt{n_j^{(i,k)}+1},
\end{equation}
with $n_j^{(i,k)}$ the number of photons from mode $j$ which are associated with the $i$th pair and position $k=L,R$ within that pair.
The equality follows from the fact that only one mode will be associated with a particular $(i,k)$, i.e. for a given $(i,k)$ there is only one $j$ for which $n_j^{(i,k)}$ is non-zero. The factor associated with a given mode $j$ is lower-bounded:
\begin{equation}
    \prod_{i=1}^H \prod_{k=L,R} \sqrt{n_j^{(i,k)}+1}\geq \sqrt{n_j+1},
\end{equation}
which occurs when $n_j^{(i,k)}$ is non-zero for only one choice of $(i,k)$. Hence the overall number of inclusion/exclusion terms is lower-bounded by
\begin{equation}
    \prod_{i=1}^H (\eta_i+1)\geq\prod_{j=1}^M \sqrt{n_j+1}.
\end{equation}

\subsection{Matching algorithm \label{edge_match}}

Since the fixed perfect matching in the eigenvalue-trace algorithm is arbitrary, we can choose it so as to create identical pairs and reduce the number of steps.
Equivalently, we can permute rows/columns in the input matrix so as to change the pairings created using the $X$ matrix.
Here, we give a greedy algorithm which chooses the pairings in the fixed perfect matching so as to minimise the number of inclusion/exclusion steps.

We start by creating a list $\vec{m}=(0,1,\dots, M-1)$. The algorithm then proceeds as follows:
\begin{enumerate}
    \item Sort $\vec{n}$ and $\vec{m}$ in descending order according to $\vec{n}$. \label{sort}
    \item If $n_1\geq 2n_2$, create $(m_1,m_1)$ pairs, which are repeated $\lfloor n_1/2 \rfloor$ times, with $\lfloor n_1/2 \rfloor$ rounding $n_1/2$ down to the nearest integer. Otherwise, create $(m_1,m_2)$ pairs which are repeated $n_2$ times.
    \item If $(m_1,m_1)$ pairs were created, subtract ${2\lfloor n_1/2 \rfloor}$ from $n_1$. Otherwise, subtract $n_2$ from $n_1$ and from $n_2$.
    \item Remove elements of $\vec{n}$ and $\vec{m}$ where $n=0$.
    \item If $\sum \vec{n} > 1$, return to step~\ref{sort}, otherwise end.
\end{enumerate}
An implementation of this algorithm can be found in the function \verb|matched_reps| of our repository~\cite{gbs_mis}.
This returns a set of pairings and a number of repeats for each pairing, $\vec{\eta}$.
Then, following the repeated pairs algorithm above, the loop hafnian can be calculated in time $\mathcal{O}\left(N^3\prod (\eta_i+1)\right)$.
This improves on eigenvalue-trace for a general matrix whenever $\vec{\eta}$ contains an entry $>1$, and this is true whenever there are at least two elements $>1$ in $\vec{n}$, or if there is at least one element $\geq 4$.

\subsection{Finite difference sieve \label{fds}}

By analogy to the Glynn formula for the permanent~\cite{Balasubramanian1980thesis, BAX1996171, GLYNN20101887}, we can find an alternative expression for the loop hafnian which uses a finite difference sieve instead of an inclusion/exclusion formula:
\begin{equation}
    \lhaf(A) = \frac{1}{2^{N/2}} \sum_{\vec{\delta}} \left( \prod_{k=1}^{N/2} \delta_k \right)  f(A X_{\vec{\delta}})
    \label{fdsieve}
\end{equation}
where $\vec{\delta}$ describes all possible $N/2$ length vectors with $\delta_i \in \{-1,1\}$. 
Here $X_{\vec{\delta}}$ is defined as:
\begin{equation}
    X_{\vec{\delta}} = \begin{pmatrix}
    \zero & \diag(\vec{\delta}) \\
    \diag(\vec{\delta}) & \zero
    \end{pmatrix}.
\end{equation}

Since an overall sign change to $\vec{\delta}$ leaves the terms inside the summation unchanged, one element of $\vec{\delta}$ can be fixed, e.g. $\delta_1=1$, and the result multiplied by 2, halving the run-time.

We can make use of repeated pairings in the finite difference sieve algorithm in the same way as above. Then, for each pairing, $\delta$ runs from $-n_{\mathrm{pair}}$ to $+n_{\mathrm{pair}}$ in steps of 2, with the different terms corresponding to how many copies of the pair are associated with a $-1$.

\begin{figure}
    \centering
    \includegraphics[width=\columnwidth]{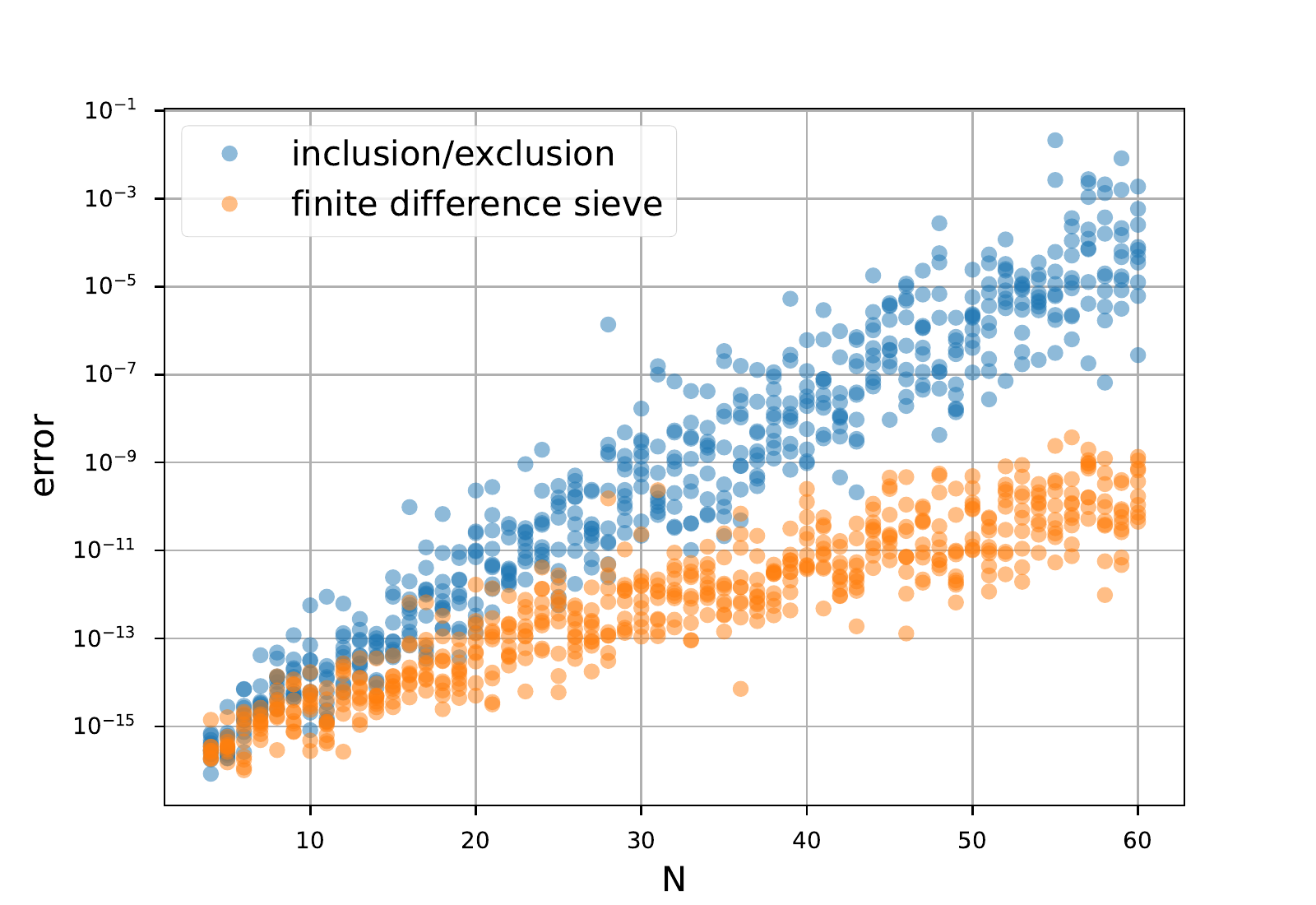}
    \caption{Numerical accuracy when comparing inclusion/exclusion and finite difference sieve based loop hafnian algorithms. We construct an $N\times N$ matrix, $C$, using 2 random $N/2 \times N/2$ matrices, $A$ and $B$, on the diagonal quadrants. We define: $\mathrm{error} = |\frac{\lhaf(C) - \lhaf(A)\lhaf(B)}{\lhaf(A)\lhaf(B)}|$. We plot the error for 10 random instances for each $N$ using 64 nodes of the Isambard HPC system.}
    \label{glynn_accuracy}
\end{figure}

We find that this method offers significant accuracy improvements over inclusion/exclusion, as shown in Fig.~\ref{glynn_accuracy}. 
In both algorithms, the absolute value of the machine precision errors accumulate at a fairly similar rate inside the sum, however in the finite difference sieve, the prefactor in equation~\ref{fdsieve} divides the value of the error by the number of terms in the sum.

\subsection{Batching probability calculations\label{batching}}
For each step of the chain-rule algorithm, we require probabilities where all but one mode has a fixed outcome, $\vec{n}_{\mathrm{fixed}}$, while one `batched' mode takes all values from 0 to the cutoff, $\ncut$. 

In the pair matching algorithm, we only input $\vec{n}_{\mathrm{fixed}}$.
If we consider the calculation when the batched mode is equal to $\ncut$, this leaves $\lfloor n_{\ncut} / 2 \rfloor$ copies of the batched mode paired to itself.
This calculation includes finding all the necessary eigenvalues required for calculating any outcome $\leq n_{\ncut}$. 
Since this is the only cubic time step within each term in the sum, we can compute all probabilities for $n \leq n_{\ncut}$ in the same time complexity as calculating $P(n_{\mathrm{fixed}}, n_{\ncut})$.
For batching across sub-detectors within the same mode in threshold detector sampling, each detector is treated independently and so has a different $\beta_i$, but this does not change the eigenvalue calculation, so these calculations can be batched in a similar way. We have implemented these methods in~\cite{gbs_mis}.

These methods could also be applied to speed up calculations of heralded non-Gaussian states in the Fock basis, as is described in ref.~\cite{quesada2019realistic}.

\subsection{Implementation details}
Our loop hafnian code is written in Python and uses Numba, a just-in-time compiler which automatically generates highly efficient code~\cite{10.1145/2833157.2833162}. 
To run efficiently on distributed systems, we use MPI for Python~\cite{DALCIN2008655}. The eigenvalue-trace algorithm is readily parallelisable as each term in the sum can be computed independently of all other terms.

Whilst testing and benchmarking our code, we ran on all major operating systems, and on x86-64 and arm64 architectures. Both Fugaku and the Isambard system (used for data in Fig.~\ref{glynn_accuracy}) use arm64 chip architectures, giving us further confidence in our run-time predictions.

\section{MIS GBS algorithms}

MIS is a Markov Chain Monte Carlo method of sampling 
which works by suggesting a state from a proposal distribution and accepting it according to the acceptance probability, Eq.~(\ref{eq:transitionprob}).
Otherwise, the previous state is added to the Markov chain.

\subsection{Independent Pairs and Singles GBS distribution \label{proposal}}
Choosing a suitable proposal distribution is an extremely important factor for MIS to be useful. 
If the proposal distribution does not match closely to the target distribution, this will result in low acceptance probabilities, and hence a very long thinning interval. 
Here, we introduce an Independent Pairs and Singles (IPS) distribution, where as the name suggests we generate multi-photon samples from many independent single-photon and pair-photon generation processes, without quantum interference between separately generated singles/pairs.
We find this is a better approximation to GBS than other efficiently simulable alternatives such as thermal states or distinguishable squeezed states.

Beginning from a pure Gaussian state which we wish to approximate, we first sample the number of individual photons created in each mode by the displacement, using Poisson distributions with the mean of the $j$th mode given by $|\alpha_j|^2$. 
We then sample the number of photon pairs created by squeezing between all mode pairs $(j,k)$ (with $j \leq k$) from a Poisson distribution with mean given by $|\bm{B}|^2_{j,k}$.
Combining all outcomes results in a photon number pattern, $\vec{n}$.

For MIS, we must calculate the probability of our generated proposal sample, $\vec{n}$.
There are many possible ways to create the same sample, corresponding to different groupings of the photons into pairs and singles.
The total probability is related to a loop hafnian, which contains a corresponding sum over all single-pair matchings.
We can write this probability as:
\begin{multline}
    Q(\vec{n}|\bm{B},\alpha) = \\
    \frac{e^{
    - \sum_j |\alpha_j|^2}
    e^{- \sum_{j,k}\frac{1}{2}|B_{j,k}|^2}}
    {\prod_i n_i!}
    \lhaf(\bm{C}_{\vec{n}}),
\end{multline}
where $\bm{C}_{\vec{n}}$ is the matrix formed by taking $|\bm{B}_{\vec{n}}|^2$ and replacing the diagonal elements with $|\vec{\alpha}_{\vec{n}}|^2$.

The loop hafnian of a positive matrix is likely to be efficient to compute approximately~\cite{rudelson2016hafnians, Gupt2019}.
However, in MIS we must also compute a loop hafnian of a complex matrix to evaluate the target probability of the sample.
Hence for convenience and simplicity we make use of the same optimised and parallelised code to compute both loop hafnians, without losing any accuracy.
This increases the run-time by at most a factor of 2.

\subsection{PNRD GBS \label{PNRD_MIS}}

We first consider the case of sampling in the photon number basis, $\vec{n}$.
We expand the sample space to include a displacement variable, $\vec{\alpha}$, so that only pure-state probabilities need to be evaluated.
The target distribution $P(\vec{n},\vec{\alpha})$ can be written as $P(\vec{\alpha})P(\vec{n}|\vec{\alpha})$ where $P(\vec{\alpha})$ is a multivariate normal distribution and hence efficient to sample from, while $P(\vec{n}|\vec{\alpha})$ is given by Eq.~(\ref{mixed_prob}) and depends on an $N\times N$ loop hafnian.
We choose $Q(\vec{\alpha})=P(\vec{\alpha})$, which results in the acceptance probability:
\begin{equation}
    p_\text{accept}=\text{min}\left(1,\frac{P(\vec{n}_i|\vec{\alpha}_i)Q(\vec{n}_{i-1}|\vec{\alpha}_{i-1})}{P(\vec{n}_{i-1}|\vec{\alpha}_{i-1})Q(\vec{n}_i|\vec{\alpha}_i)}\right),
\end{equation}
so the acceptance probability does not depend on the probability density of $\vec{\alpha}$.

In some cases it may be useful to fix the total photon number $N$ when sampling; for example verification methods often focus on samples of a particular $N$.
In MIS it is possible to fix $N$ by post-selecting our proposed states - this does not add appreciably to the run-time, since generating proposed states can be done efficiently and the computational effort is dominated by calculating $p_\text{accept}$.
In this case, the acceptance probability is
\begin{align}
p_\text{accept}&=\text{min}\left(1,\frac{P(\vec{n}_i,\vec{\alpha}_i|N)Q(\vec{n}_{i-1},\vec{\alpha}_{i-1}|N)}{P(\vec{n}_{i-1},\vec{\alpha}_{i-1}|N)Q(\vec{n}_i,\vec{\alpha}_i|N)}\right) \nonumber \\
&=\text{min}\left(1,\frac{P(\vec{n}_i,\vec{\alpha}_i,N)Q(\vec{n}_{i-1},\vec{\alpha}_{i-1},N)}{P(\vec{n}_{i-1},\vec{\alpha}_{i-1},N)Q(\vec{n}_i,\vec{\alpha}_i,N)}\right) \nonumber \\
&=\text{min}\left(1,\frac{P(\vec{n}_i,\vec{\alpha}_i)Q(\vec{n}_{i-1},\vec{\alpha}_{i-1})}{P(\vec{n}_{i-1},\vec{\alpha}_{i-1})Q(\vec{n}_i,\vec{\alpha}_i)}\right),
\end{align}
where in the second line we used the definition of conditional probability $P(\vec{n}_i,\vec{\alpha}_i,N)=P(\vec{n}_i,\vec{\alpha}_i|N)P(N)$, and the $P(N)$'s cancel and so do the $Q(N)$'s.
In the third line, we know that if we are post-selecting, all $\vec{n}$ will automatically satisfy $N$ so it is a redundant variable.
Hence an identical $p_\text{accept}$ can be used when fixing $N$.

We outline the algorithm below. To sample from a state with vector of means $\bm{R}$ and covariance matrix $\bm{V}$:

\begin{enumerate}
    \item Use the Williamson decomposition to write $\bm{V}=\bm{T}+\bm{W}$, where $\bm{T}$ is the covariance matrix of a pure state. Calculate the matrix $\B$ based on $\bm{T}$.
    \item Sample a displacement vector $\bm{R'}$ from the multivariate normal distribution $\bm{R'} \sim \mathcal{N}(\bm{R},\bm{W})$. Calculate the complex displacement $\vec{\alpha}'_1$ from $\bm{R'}$.
    \item Sample a photon pattern $\vec{n}_1$ from $Q(\vec{n}|\B,\vec{\alpha}_1)$.
    This involves sampling from Poissonian distributions.
    \item Start a Markov chain from the state $(\vec{n}_1,\vec{\alpha}'_1)$.
    \item For step $i$ in the Markov chain from 2 to the desired length:
    \begin{enumerate}
        \item Sample a new displacement vector $\vec{\alpha}'_i$.
        \item Sample a new photon pattern $\vec{n}_i$ from $Q(\vec{n})$ for the pure state with displacement $\vec{\alpha}'_i$ and covariance matrix $\bm{T}$.
        \item Calculate the acceptance probability $p_{\text{accept}}$.
        \item Add $(\vec{n}_i,\vec{\alpha}'_i)$ to the chain with probability $p_{\text{accept}}$, otherwise add the previous state again.
    \end{enumerate}
    \item Keep only the $\vec{n}$ values in the chain (ignore $\vec{\alpha}'$). Discard the first $\tburn$ samples and then keep every 1 in $\tthin$ samples.
    
\end{enumerate}

\subsection{Threshold detector GBS \label{click_mis}}

For MIS with threshold detectors, we also need to include the fan out of each mode into sub-detectors, where we only register the position of the `first' photon.
So we now expand the sample space to include a variable describing this position, $x$.
We take the limit of a large number of sub-detectors where $x$ becomes a continuous variable, and choose larger $x$ to correspond to `earlier' detections.
 
The POVM element for a click outcome where the first photon is at position $x$ can be written:
\begin{equation}
    \pi_c(x)=\sum_{n=1}^\infty p(x|n)\ket{n}\bra{n},
    \label{pi_cx}
\end{equation}
with $p(x|n)=nx^{n-1}$. $\pi_c(x)$ is closely related to the POVM element for measuring a single photon after a loss of $x$:
\begin{equation}
    \Pi_L(x)=\sum_{j=1}^\infty j(1-x)x^{j-1}\ket{j}\bra{j}=(1-x)\pi_c(x).
    \label{eq:clickx}
\end{equation}
Hence we can express the probability of a click pattern $\vec{c}$ with an accompanying $\vec{x}$ in terms of the probability of obtaining the same pattern of single photons, but from a covariance matrix $\bm{V}(\vec{x})$ where the loss $x_m$ has been applied to the $m$th mode (for unoccupied modes $x_m=0$):
\begin{equation}
    P_c(\vec{c},\vec{x},\vec{\alpha}')=\frac{P(\vec{\alpha}') P_n(\vec{c}|\vec{x},\vec{\alpha}')}{\prod_m(1-x_m)},
\end{equation}
where as before $\vec{\alpha}'$ is a complex displacement vector chosen from a multivariate normal distribution.
Since $\bm{V}(\vec{x})$ is a mixed state, we expand it as an ensemble of pure states with differing displacement vectors $\vec{\alpha}''$:
\begin{equation}
    P_c(\vec{c},\vec{x},\vec{\alpha}',\vec{\alpha}'')=\frac{P(\vec{\alpha}')P(\vec{\alpha}''|\vec{x},\vec{\alpha}')P_n(\vec{c}|\vec{x},\vec{\alpha}'')}{\prod_m(1-x_m)},
    \label{thresh_targ}
\end{equation}
where $P(\vec{\alpha}''|\vec{x},\vec{\alpha}')$ is the probability distribution of $\vec{\alpha}''$, depending on the applied loss, $\vec{x}$, and the complex displacement before the loss, $\vec{\alpha}'$.
$P_n(\vec{c}|\vec{x},\vec{\alpha}'')$ is the photon number pattern probability of a pure state and can be calculated with an $N_c\times N_c$ loop hafnian, in time $\mathcal{O}(N_c^3 2^{N_c/2})$, resulting in a quadratic speedup compared to a Torontonian.
If we sample $(\vec{c},\vec{x},\vec{\alpha}'',\vec{\alpha}')$ and then ignore the $\vec{x}$ and $\vec{\alpha}$ outcomes, this is equivalent to sampling from $P_c(\vec{c})$ as desired. 

To generate proposal samples, we begin by generating a displacement vector  $\vec{\alpha}'$ and photon number pattern $\vec{n}$ as in Appendix~\ref{PNRD_MIS}.
Then a $\vec{x}$ vector can be generated by sampling from $p(x|n)$ for each element, and a click pattern $\vec{c}$ taken by reducing each $>0$ element of $\vec{n}$ to a 1.
The loss $\vec{x}$ is applied to the state, resulting in an updated displacement $\vec{\alpha}'(x)$ and covariance matrix $\bm{V}(\vec{x})$, from which a Williamson decomposition can be used to sample a pure state - with a displacement $\vec{\alpha}''$ and covariance matrix $T'$.

The proposal probability, marginalised over $\vec{n}$, can be written
\begin{equation}
    Q_c(\vec{c},\vec{x},\vec{\alpha}',\vec{\alpha}'')=P(\vec{\alpha})P(\vec{\alpha}''|\vec{x},\vec{\alpha}')Q_c(\vec{c},\vec{x}|\vec{\alpha}'),
    \label{thresh_prop}
\end{equation}
where we note that the proposal distribution for $(\vec{c}$, $\vec{x})$ is conditioned on $\vec{\alpha}'$ rather than $\vec{\alpha}''$, which is the last variable to be chosen.
As with the target distribution, this probability can be rewritten in terms of a pattern of single photons after application of a loss:
\begin{equation}
    Q_c(\vec{c},\vec{x}|\vec{\alpha}')=\frac{Q_n(\vec{c}|\vec{x},\vec{\alpha}')}{\prod_m (1-x_m)}.
\end{equation}
The probability of detecting a pattern of single photons from IPS after loss is still given by the loop hafnian of a non-negative matrix:
\begin{equation}
    Q_n(\vec{c}|\vec{x},\vec{\alpha}')\propto \lhaf(\bm{C}_{\vec{c}}(\vec{x})),
\end{equation}
where we take
\begin{equation}
    \bm{C}_{j,k}(\vec{x})=(1-x_j)(1-x_k)|B_{j,k}|^2,
\end{equation}
except for diagonal elements
\begin{equation}
     \bm{C}_{j,j}(\vec{x})=(1-x_j)\left(|\alpha_j|^2+\sum_k x_k |B_{j,k}|^2\right),
\end{equation}
and form $\bm{C}_{\vec{c}}(\vec{x})$ by keeping the elements of $\bm{C}$ where $c=1$.
This results in an acceptance probability:
\begin{equation}
    p_\text{accept}=\text{min}\left(1,\frac{P_n(\vec{c}_i|\vec{x}_i,\vec{\alpha}''_i)Q_n(\vec{c}_{i-1}|\vec{x}_{i-1},\vec{\alpha}'_{i-1})}{Q_n(\vec{c}_i|\vec{x}_i,\vec{\alpha}'_i)P_n(\vec{c}_{i-1}|\vec{x}_{i-1},\vec{\alpha}''_{i-1})}\right).
    \label{thresh_acc}
\end{equation}
We outline the steps of the MIS algorithm below.
\begin{enumerate}
     \item Use the Williamson decomposition to write $\bm{V}=\bm{T}+\bm{W}$, where $\bm{T}$ is the covariance matrix of a pure state.
     \item Sample the starting state from the proposal distribution
     \begin{enumerate}
    \item Sample a complex displacement vector $\vec{\alpha}'_1$.
    \item Sample a photon pattern $\vec{n}_1$ from $Q(\vec{n}|\vec{\alpha}')$.
    Find $\vec{c}_1$ from $\vec{n}_1$ by fixing all $n_i>1$ as $c_i=1$.
    \item If post-selecting on the number of clicks, repeat the above steps until $\vec{c}$ contains the desired number of clicks.
    \item Sample the loss, $\vec{x}_1$, conditional on the photon number pattern $\vec{n}_1$ using $p(x|n)$.
    \item Apply the loss $\vec{x}_1$ to the displacement vector $\vec{\alpha}'_1$  and the covariance matrix $\bm{T}$, resulting in $\vec{\alpha}'_1(\vec{x})$ and $\bm{V}(\vec{x})$.
    \item Perform a Williamson decomposition on the mixed state to obtain a pure state covariance matrix $\bm{T'}$ and sample a new complex displacement vector $\vec{\alpha}_1''$.
    \end{enumerate}
    \item Start the Markov chain from the state $(\vec{c}_1,\vec{x}_1,\vec{\alpha}'_1,\vec{\alpha}''_1)$. Calculate the target probability using Eq.~\ref{thresh_targ} and the proposal probability using Eq.~\ref{thresh_prop}.
    \item For step i in the Markov chain from 2 to the desired length:
    \begin{enumerate}
        \item Sample another proposal sample $(\vec{c}_i,\vec{x}_i,\vec{\alpha}'_i,\vec{\alpha}''_i)$.
        \item Calculate the target and proposal probabilities for this state.
        \item Calculate the acceptance probability $p_\text{accept}$ using Eq.~\ref{thresh_acc}.
        \item Add $(\vec{c}_i,\vec{x}_i,\vec{\alpha}'_i,\vec{\alpha}''_i)$ to the chain with probability $p_\text{accept}$, otherwise add the previous state again.
    \end{enumerate}
    \item Keep only the $\vec{c}$ values in the chain (ignore $\vec{x}$, $\vec{\alpha}'$ and $\vec{\alpha}''$). Discard the first $\tburn$ samples and then keep every 1 in $\tthin$ samples.
\end{enumerate}

\subsection{Thinning interval and burn-in time scaling \label{mis_scaling}}
To investigate the scaling of our algorithms, 
we fix the number of photons to the mean photon number number, rounded to the nearest integer.
The tests described in this section are applicable to both PNRD and threshold GBS unless stated otherwise. 
However, we only implement them for the number resolving case.

\begin{figure}
    \centering
    \includegraphics[width=0.5\textwidth]{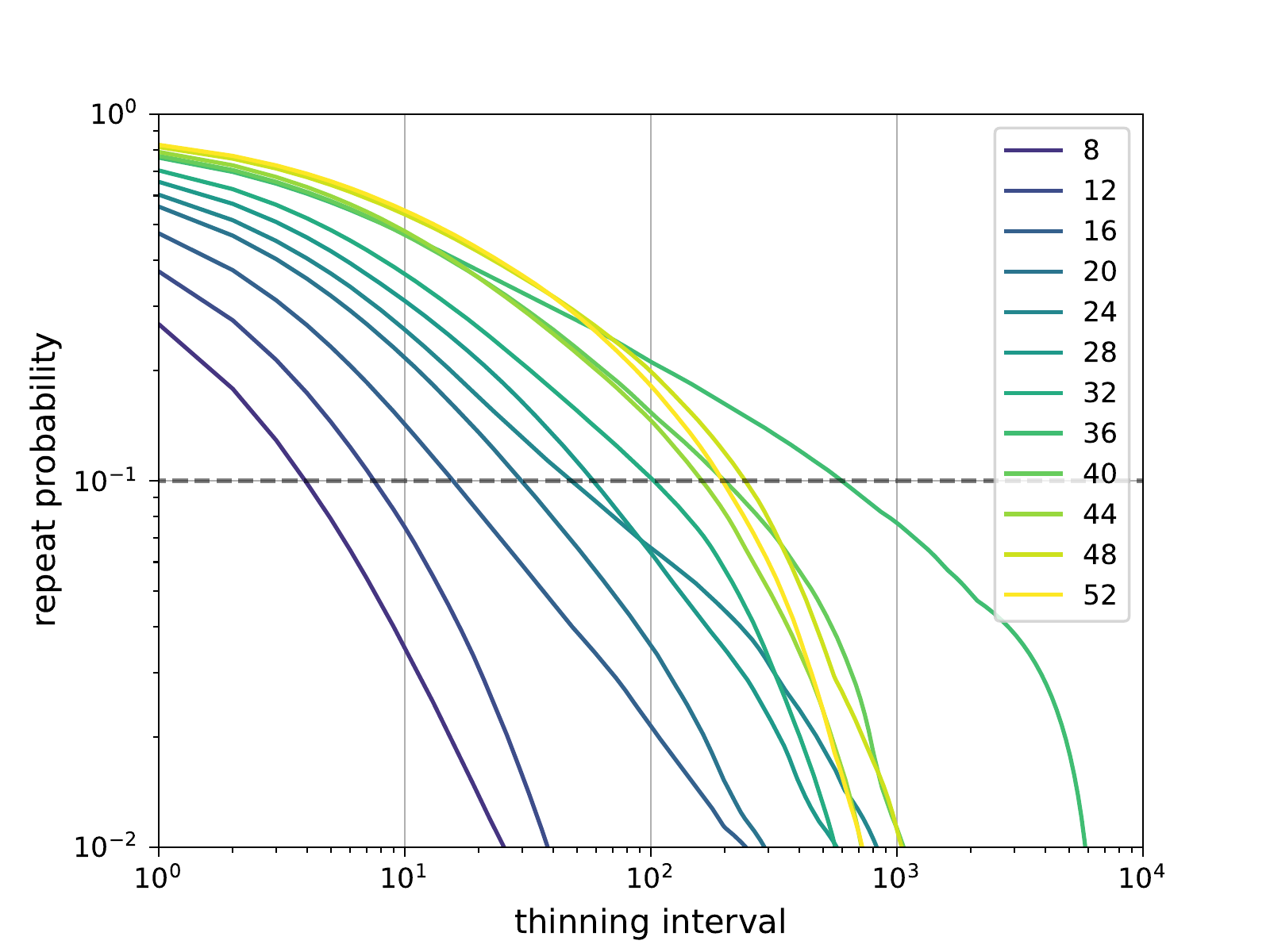}
    \caption{Estimated probability of repeated samples. Calculated for each $M$ from 10 Haar random unitary matrices, with a 10,000 long MIS chain for each unitary.}
    \label{repeats_probs}
\end{figure}

\begin{figure}
    \centering
    \includegraphics[width=0.5\textwidth]{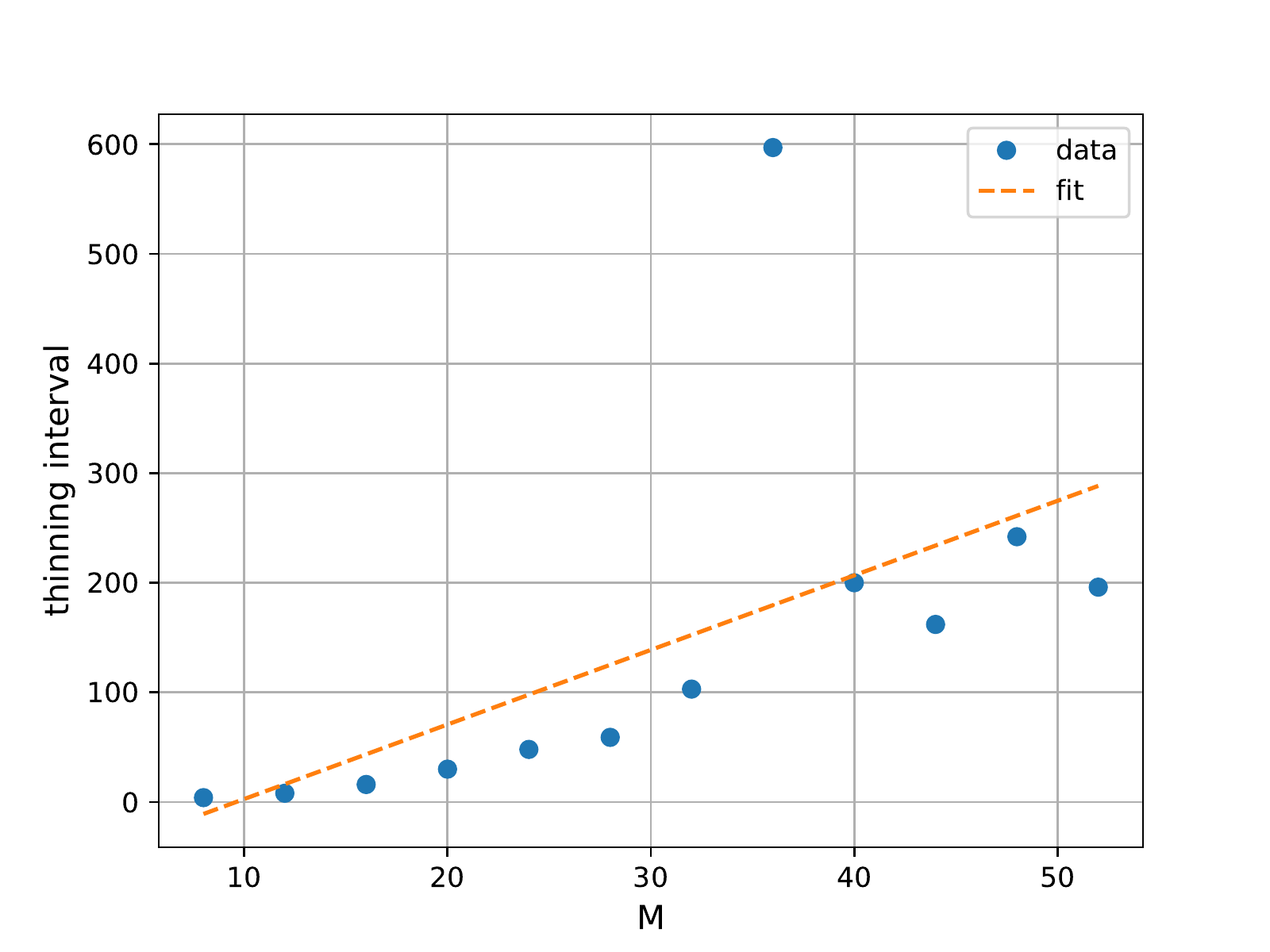}
    \caption{Thinning intervals generated from the data in Fig.~\ref{repeats_probs}, by choosing 0.1 as an acceptable repeat probability. The $M=36$ data point is far from the line of best fit, likely due to a sample with an anomalously large separation between target and proposal probabilities.
    The linear fit follows $\tau_{thin} = 6.81M - 65.3$.}
    \label{thinning_rate_fit}
\end{figure}

It is important to be able to predict the run-time of simulations before they are performed. 
For MIS methods, this is challenging as thinning intervals and burn-in times depend on how close the proposal distribution is to the target distribution. 
To construct heuristics to allow us to make these predictions, we investigate how the thinning interval and burn-in time scale with the number of modes $M$.
However, we wish to highlight that the requirements on accuracy and sample autocorrelation will vary depending on what is desired from the simulation. 
Therefore the results in this section should be viewed as a guide for how to predict the scaling, rather than as a prescriptive guide for what parameters should be used.

To predict the thinning interval, $\tthin$, we investigate systems of different sizes with $M$ varied between 8 and 52 in steps of 4.
For each $M$ we choose 10 Haar random interferometers, and implement an MIS chain with 10,000 steps.
In Fig.~\ref{repeats_probs}, we plot the estimated probability of a sample being repeated as a function of the thinning interval.
From this we extract the thinning interval required to suppress the repeat probability to 0.1 for each $M$ and perform a linear fit on this data,
shown in Fig.~\ref{thinning_rate_fit}.

The data for $M=36$ appears anomalous.
We believe this is caused by one of the chains drawing a proposal sample which has an unusually large target/proposal probability ratio.
Such events can cause chains to reject a very large number of samples before accepting a new proposal sample. 
The large degree of autocorrelation which is created by events such as this are an intrinsic drawback of the MIS method, and so we do not discard this data.

The second parameter we need to determine is the burn-in time, $\tau_\text{burn}$.
We know that the chain begins sampling from the proposal distribution and over time converges to the target distribution.
It will converge continuously, getting asymptotically closer to the target distribution, but at some point it will be close enough that the change will not be noticeable from finite sample sizes.
Therefore, we can analyse when our distribution appears to be stationary.
We provide two tests to predict how the burn-in time scales with the number of modes.

For the first, we use a Bayesian likelihood ratio test for each burn-in time until we see no improvement.
The likelihood ratio tests whether a set of samples $s$ is more likely to have come from the target distribution or an adversary distribution. 
To test how close our distribution is to the target, $\mathcal{P}$, or proposal, $\mathcal{Q}$, we choose the proposal distribution as the adversary.
We begin with the ratio
\begin{equation}
    \chi=\frac{p(\mathcal{P}|s)}{p(\mathcal{Q}|s)}=\frac{p(s|\mathcal{P})p(\mathcal{P})p(s)}{p(s|\mathcal{Q})p(\mathcal{Q})p(s)}.
\end{equation}
If we assume equal priors $p(\mathcal{P})=p(\mathcal{Q})$, this simplifies to 
\begin{equation}
    \chi=\frac{p(s|\mathcal{P})}{p(s|\mathcal{Q})}.
\end{equation}
Assuming that the probability distribution is either $\mathcal{P}$ or $\mathcal{Q}$ and so $p(\mathcal{P}|s)+p(\mathcal{Q}|s)=1$, we can write
\begin{align}
    p(\mathcal{P}|s)&=p(\mathcal{Q}|s) \chi = (1-p(\mathcal{P}|s)) \chi \\
    \implies p(\mathcal{P}|s)&=\frac{\chi}{1+\chi}.
\end{align}

Here our samples $s_i$ are described by $(\vec{\alpha}_i,\vec{n}_i)$ which we sample from the chain. So we can write $p(s_i|\mathcal{P})= p(\vec{\alpha}_i,\vec{n}_i|\mathcal{P})= p(\vec{\alpha}_i|\mathcal{P})p(\vec{n}_i|\vec{\alpha}_i,\mathcal{P})$. For purposes of benchmarking the efficiency, we fix the number of photons and so we have to adjust for post-selecting on $N$ photons. We still assume equal priors, now $p(\mathcal{P}|N)=p(\mathcal{Q}|N)$. So the likelihood ratio becomes

\begin{align}
    \chi&=\frac{p(s|\mathcal{P},N)}{p(s|\mathcal{Q},N)}=\prod_i\frac{P(\vec{\alpha}_i,\vec{n}_i|N)}{Q(\vec{\alpha}_i,\vec{n}_i|N)}\\
    &=\prod_i \frac{P(\vec{n}_i,N|\vec{\alpha}_i)Q(N)}{Q(\vec{n}_i,N|\vec{\alpha}_i)P(N)},
\end{align}
where we use the fact that $P(\vec{\alpha}_i)=Q(\vec{\alpha}_i)$.
This only requires the calculation of pure-state probabilities and the probability of getting $N$ photons in both the proposal and target distribution. This can be done for PNRDs with no additional cost to the sampling algorithm as we must calculate the pure-state probabilities in the formation of our chain, and calculating the probabilities of $N$ photons is efficient. However, for threshold detectors, although we could add the $x$ variable, we are not aware of a way to calculate the probability of $N_c$ clicks for either distribution. Therefore we do not apply this test to threshold detectors. 

\begin{figure}
    \centering
    \includegraphics[width=\columnwidth]{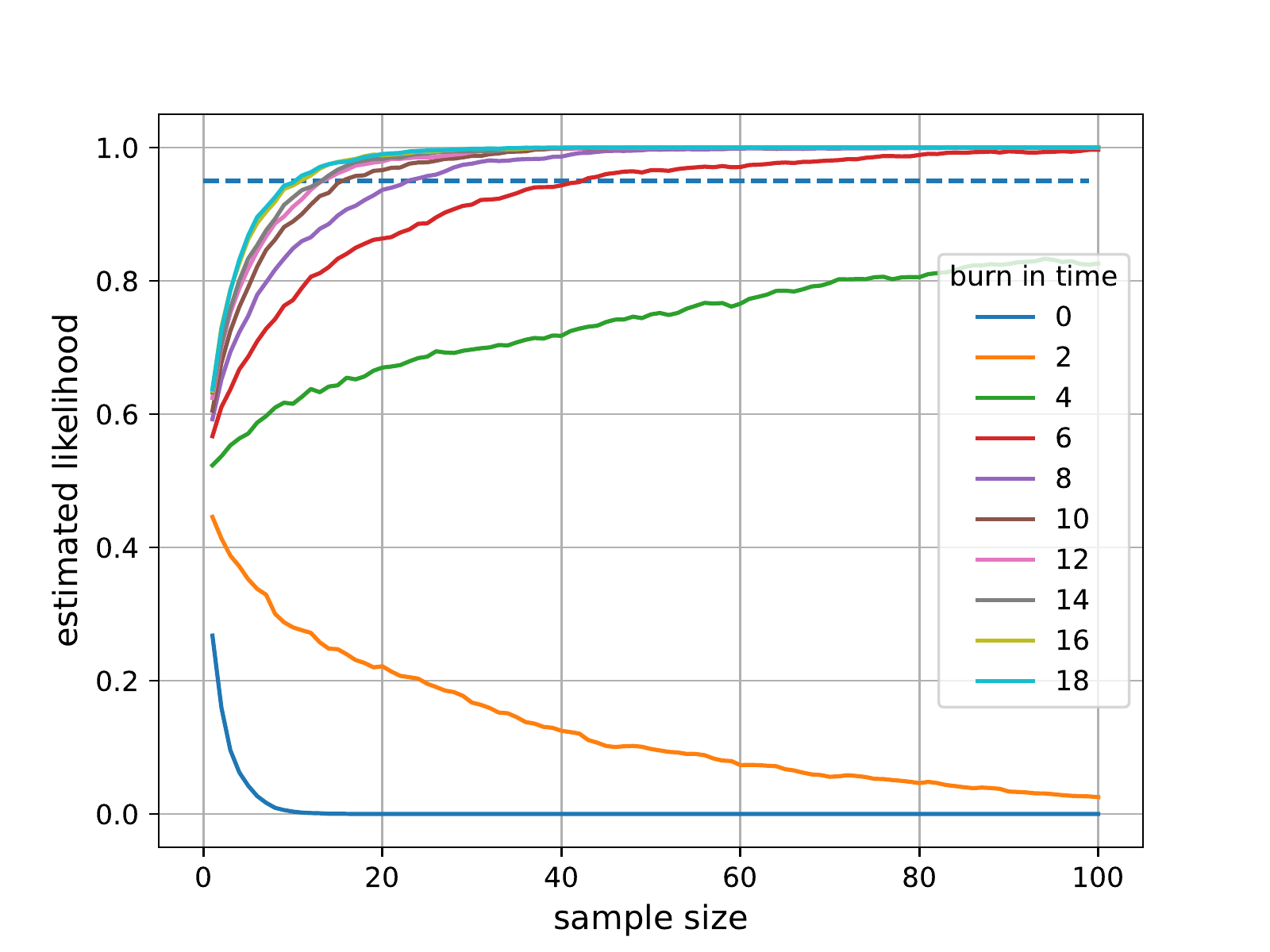}
    \caption{The estimated likelihood ratio as a function of the number of samples included for a range of burn-in times. The plot shows how the likelihood converges to either 0 if the test fails or 1 if it passes for increasing burn-in times averaged over 1000 Haar random unitaries in 28 modes. We wish to find the sample size at which the likelihood ratio reaches 0.95 as indicated by the dashed line.}
    \label{likelihood_samplesize}
\end{figure}

\begin{figure}
    \centering
    \includegraphics[width=\columnwidth]{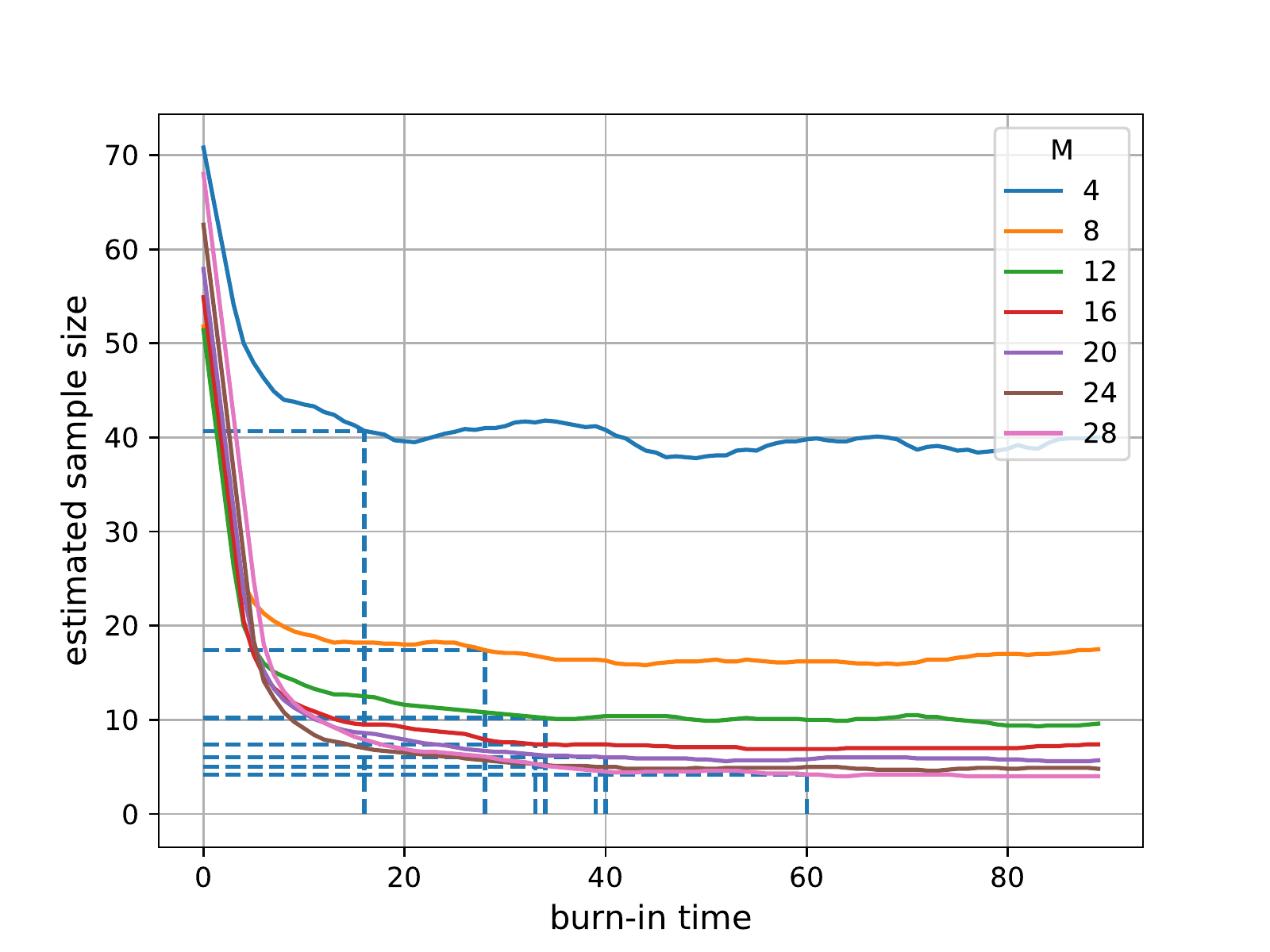}
    \caption{The estimated sample size required to give a likelihood ratio of 0.95 for burn-in times between 0 and 100 (averaging over 10 burn-in times). We wish to find the burn-in time beyond which we see no improvement in the number of samples required, as indicated by the dashed lines.}
    \label{samplesize_burnin}
\end{figure}

\begin{figure}
    \centering
    \includegraphics[width=\columnwidth]{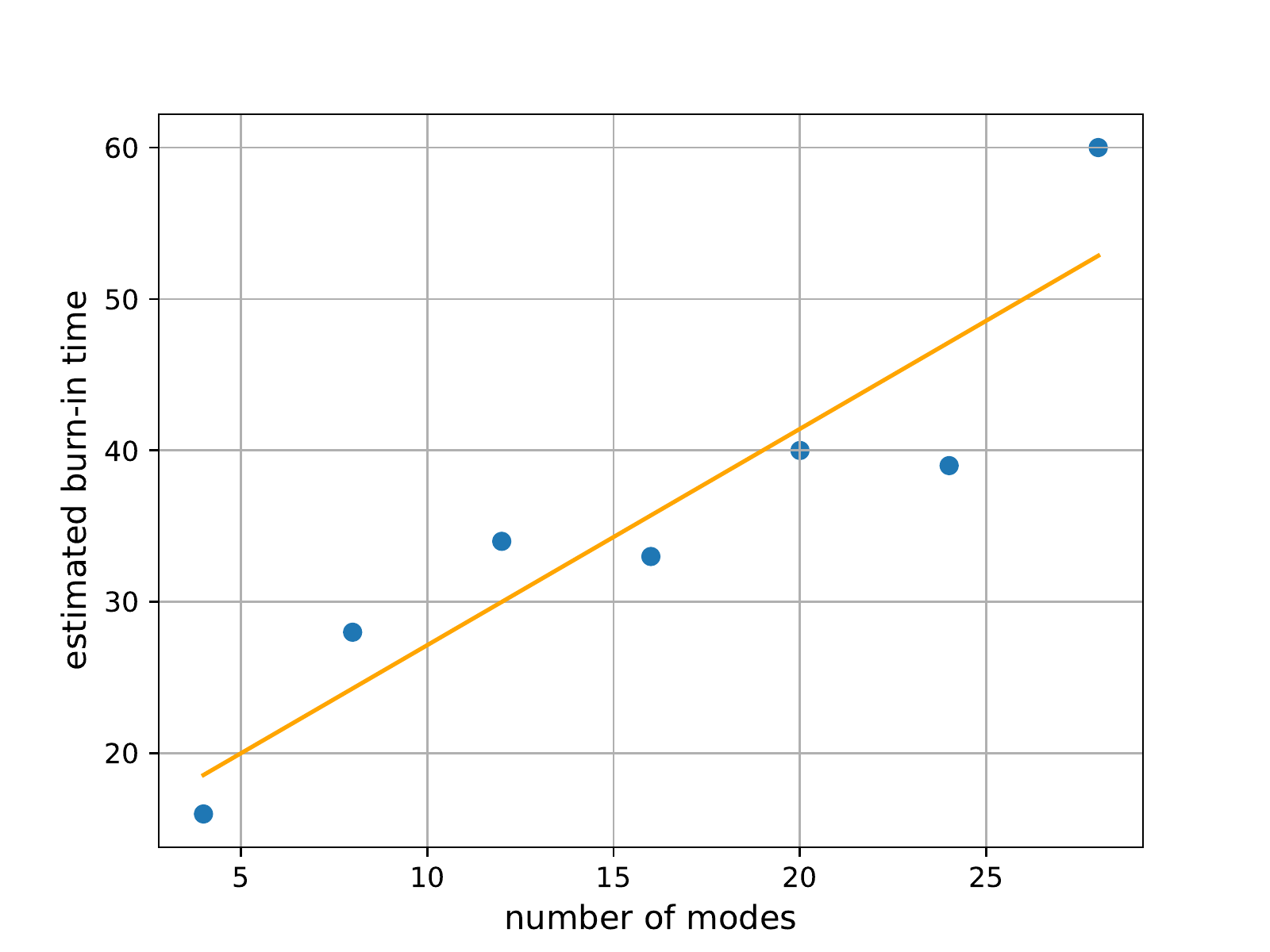}
    \caption{The estimated burn-in time from the likelihood test as a function of the number of modes. The likelihood is calculated for increasing sample size for up to a burn-in time of 100. We estimate at which burn-in time we do not see an improvement to the likelihood. We find the sample size required to reach a likelihood of 0.95 and find the burn-in at which it is within 5\% of the final value. Each likelihood was estimated by averaging over 1000 Haar random unitaries. We use a linear fit of the data to predict a relationship of $\tau_\text{burn}=1.43M+12.86$. }
    \label{convergence1}
\end{figure}

We evaluate this probability for all burn-in times up to 100, for an increasing sample size up to 100. As we increase the sample size, the likelihood should eventually converge to either 0 (if it fails) or 1 (if it passes) - see Fig.~\ref{likelihood_samplesize}. The closer the sampled distribution is to the target, the faster it converges to 1, so we can test how close our distribution is for different burn-in times by sampling and comparing the rate of the convergence to 1.
To isolate the burn-in for testing, we need to start a new chain every time we sample.
As a metric for comparing the rates of convergence, we find the sample size required to reach a likelihood of 0.95.
As we increase the burn-in time, the sample size should decrease and approach the minimum.
We find the burn-in time at which the sample size is within 5\% of the estimated minimum. Each likelihood is estimated by averaging over 1000 Haar random unitaries.
Despite this, the likelihood still gives quite noisy data and we further average over a range of 10 burn-in times, ie the likelihood at burn-in $i$ is given by the average of the likelihood for burn-in times between $i$ and $i+9$  (see Fig.~\ref{samplesize_burnin}).
The minimum is estimated in a similar way where we average over the last 20 burn-in times, when we can assume it has converged.
Fig.~\ref{convergence1} shows the estimated burn-in time for up to 28 modes and we extrapolate the linear fit to give an estimate of $\tau_\text{burn}=155$ for 100 modes.

\begin{figure}
    \centering
    \includegraphics[width=\columnwidth]{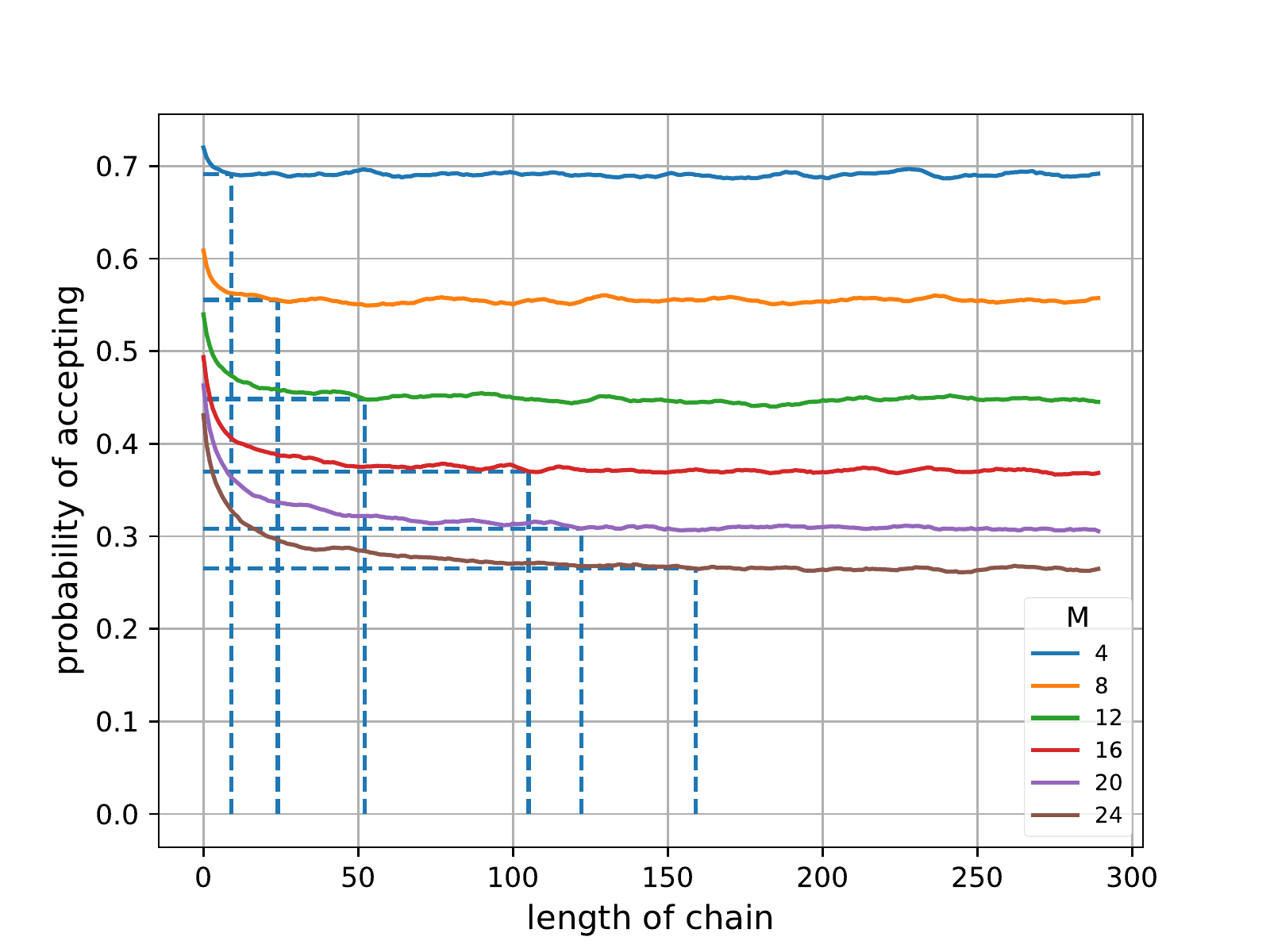}
    \caption{The acceptance rate as a function of the length of the chain for various values of $M$, the number of modes. We find the point in the chain at which the acceptance rate becomes approximately constant. This plot shows an example for one Haar random unitary for each $M$. The dashed horizontal lines show when the curve reaches withing 0.001 of the estimated final value.}
    \label{acceptancerate}
\end{figure}

\begin{figure}
    \centering
    \includegraphics[width=\columnwidth]{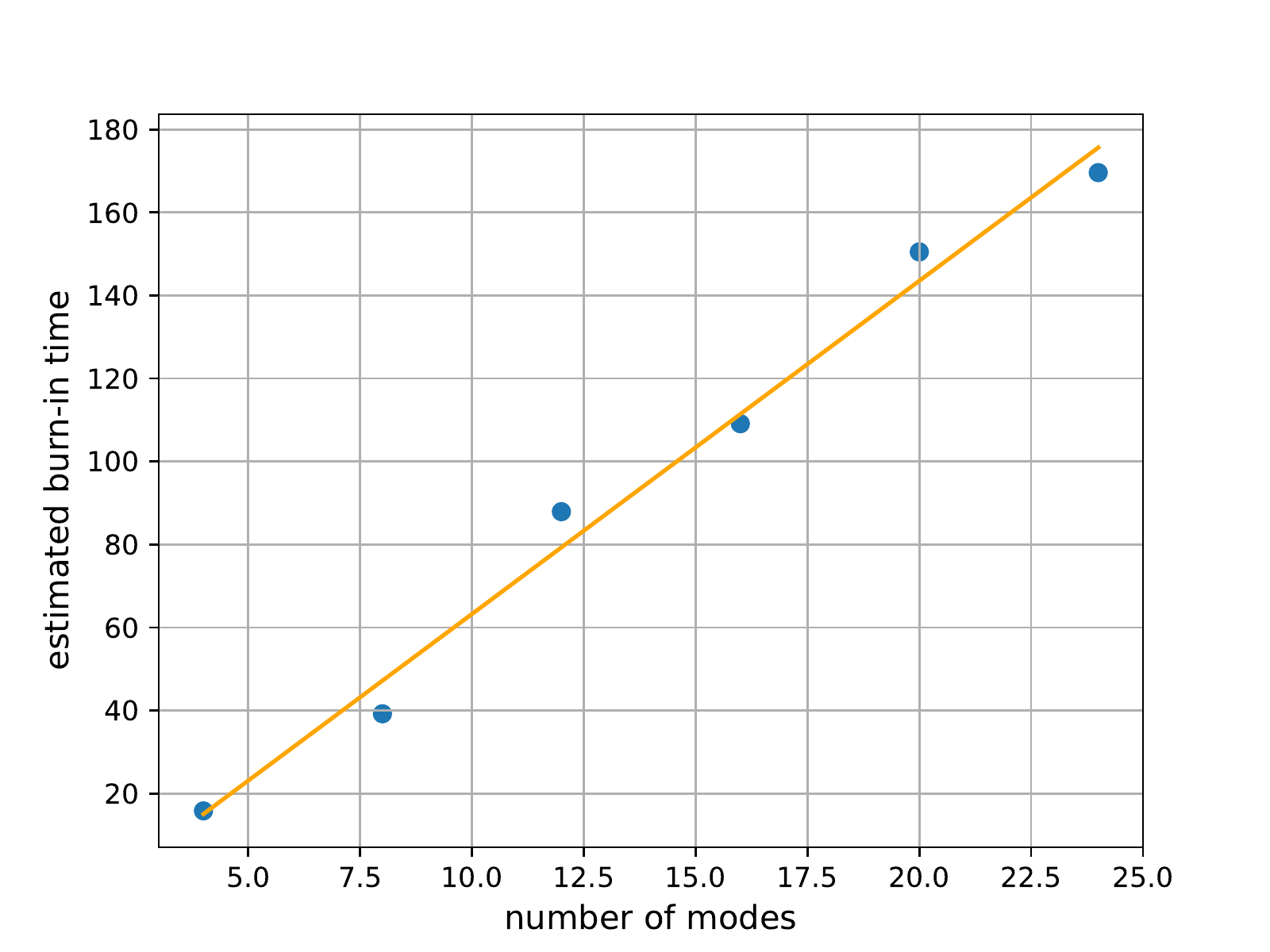}
    \caption{The estimated burn-in time from the acceptance rate test as a function of the number of modes. We estimate the burn-in beyond which the probability of accepting the proposed sample is approximately constant. Each burn-in was estimated by averaging over 10 Haar random unitaries. We use a linear fit of the data to predict a relationship of $\tau_\text{burn}=8.03M-17.06$. }
    \label{convergence2}
\end{figure}

For the second test, we note that the rate of accepting a proposed sample decreases towards an asymptotic minimum value as the chain converges.
This minimum value would be reached only when sampling from the target distribution.
We estimate the probability of accepting at each burn-in time up to 300 by running 10,000 chains and counting the number of times we accept for a Haar random unitary.
As with the likelihood test, we still have noisy data and smooth out the curve by averaging across 10 burn-in times.
We choose the burn-in time at which the probability of accepting is no more than 0.001 greater than the estimated minimum value.
Again we estimate the minimum value from the end of our chain, averaging over the last 50 burn-in times.
As long as the estimated burn-in time is significantly before the end of the chain, we can be reassured that the probability of accepting is changing slowly enough to consider the chain to have converged by the maximum burn-in time we test.
 See Fig.~\ref{acceptancerate} for an example of how the acceptance rate varies with the chain length.
We run this test for 10 Haar random unitaries and find the average burn-in time for each $M$ up to 24, shown in Fig.~\ref{convergence2}, which extrapolating gives an estimate of $\tau_\text{burn}=785$ for 100 modes.

We note that these two tests give significantly different estimates for the burn-in times.
This is likely due to two reasons.
The first is that the likelihood may be less sensitive to the convergence and doesn't distinguish between two close distributions as well.
The data for this test is more noisy and so that may hide small differences in the distributions.
The second is that we fix it to have converged further for the acceptance rate test.
It is an arbitrary choice to decide how close you require the distribution to be to the target distribution.
In both tests, we are limited by how noisy our data is from finite sampling.
If our acceptance rate test is less noisy, we are able to find the burn-in times for a better convergence.
The important findings from our numerical analysis are that the burn-in time seems to scale approximately linearly with the number of modes and the gradient of the scaling depends on how close you want the distribution you are sampling from to be to the target distribution.

\begin{figure}[t]
    \centering
    \includegraphics[width=0.475\textwidth]{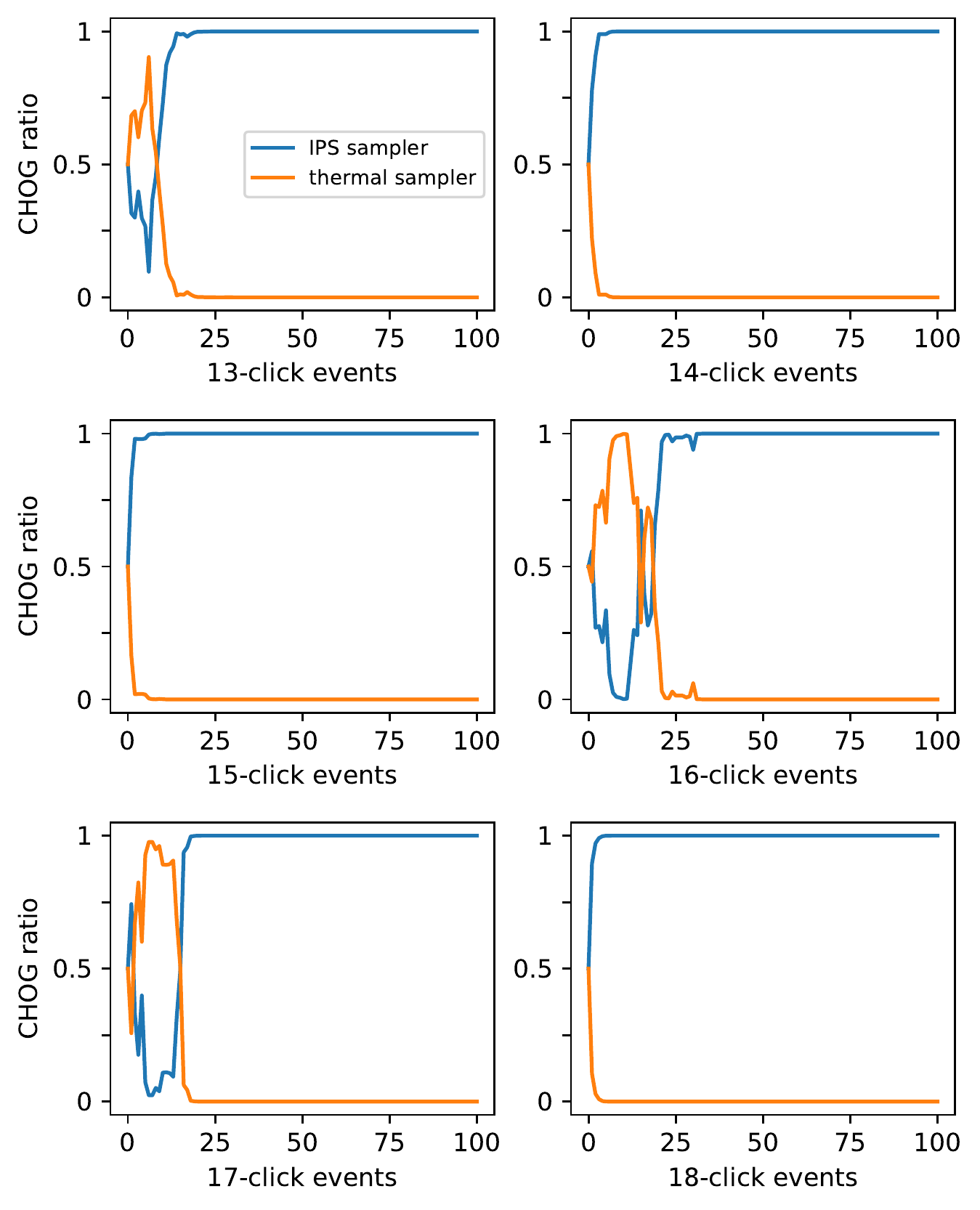}
    \caption{$M=52$ CHOG tests for different total click numbers to determine the relative likelihood of IPS and thermal samples from the ideal click distribution. Convergence to 1 for the IPS samplers indicates they are more likely to have come from the click distribution than the thermal samples.}
    \label{chog_ips_thermal}
\end{figure}

\section{Validation tests \label{verification}}
\subsection{CHOG ratio}
As we approach a scale where exact validation of samples becomes unfeasible, we can use the Chen Heavy Output Generation (CHOG) ratio test outlined in ref.~\cite{Zhong1460}. Certain output patterns from a random optical network occur more frequently due to constructive interference and it is thought to be difficult to replicate this observation with classical samplers. This adversarial test assesses the relative likelihood of two sets of samples (`trial' and `adversarial') being drawn from a given ideal distribution. As samples are drawn, the CHOG ratio is updated:

\begin{align}
    r_{\mathrm{CHOG}}&=\frac{P_{\mathrm{ideal}}(\mathrm{samples}_{\mathrm{trial}})}{P_{\mathrm{ideal}}(\mathrm{samples}_{\mathrm{trial}})+P_{\mathrm{ideal}}(\mathrm{samples}_{\mathrm{adv}})}\\
    &=\left(1+\prod_{j}\frac{P_{\mathrm{ideal}}(\mathrm{sample}_{\mathrm{adv}}(j))}{P_{\mathrm{ideal}}(\mathrm{sample}_{\mathrm{trial}}(j))}\right)^{-1}.\label{eq:chog}
\end{align}

Convergence to a value of 1 indicates that the trial samples were more likely to be drawn from the ideal distribution than the adversarial samples. In the case of click detection samples, the probabilities of observing these samples from the ideal (squeezed) distribution are calculated using the Torontonian.

In the work by USTC~\cite{Zhong1460}, click samples are drawn from the trial GBS experiment and an adversarial thermal sampler. A value of 1 indicates that the GBS samples were more likely to be drawn from the ideal distribution than the thermal samples. For Jiŭzhāng, GBS samples corresponding to fixed total numbers of clicks of between 26 and 38 were validated against a thermal sampler.

\subsubsection{IPS vs thermal}
\label{ips_vs_thermal}

Our IPS sampler (used as a proposal distribution for the MIS algorithm) naturally incorporates constructive interference of pairs of photons. Here, we use it to generate trial samples and apply the CHOG test to validate against an adversarial thermal sampler. For the IPS sampler we use a covariance matrix corresponding to 13 sources of two-mode squeezed vacuum, with squeezing parameter $r=1.55$ and transmission $\eta=0.3$, injected into a Haar random 52-mode unitary interferometer. The covariance matrix for the thermal sampler is constructed using the same transmissions and unitary interferometer, but now injected with 26 thermal states, each with mean photon number $n_{th}=\sinh^{2}(r)$. We update the CHOG ratio using equation~\ref{eq:chog} and results for click numbers of between 13 and 18 are shown in Fig.~\ref{chog_ips_thermal}. The convergence to 1 for the IPS sampler shows that those samples are more likely to have been drawn from the ideal distribution than the thermal samples.

Hence, we have shown that our IPS sampler -- from which samples can be efficiently drawn classically -- passes the CHOG test against a thermal sampler in a similar way to the experimental GBS samples from Jiŭzhāng.
This challenges the usefulness of the CHOG test against a thermal adversarial sampler in validating quantum computational complexity of GBS.
It also suggests that the IPS distribution should be used as an adversary model to test against in future experiments.
Because the IPS distribution contains no interference between different photon pairs, it could be considered as the distribution generated by squeezers with zero spectral purity.
Following this intuition, we also suggest a finite purity adversary (i.e. squeezing across $\ge 2$ Schmidt modes) as another important, more challenging, model to test against.
Calculating probabilities of Gaussian states in the presence of spectral impurity has been investigated in ref.~\cite{thomas2020general}.

\begin{figure}[t]
    \centering
    \includegraphics[width=0.475\textwidth]{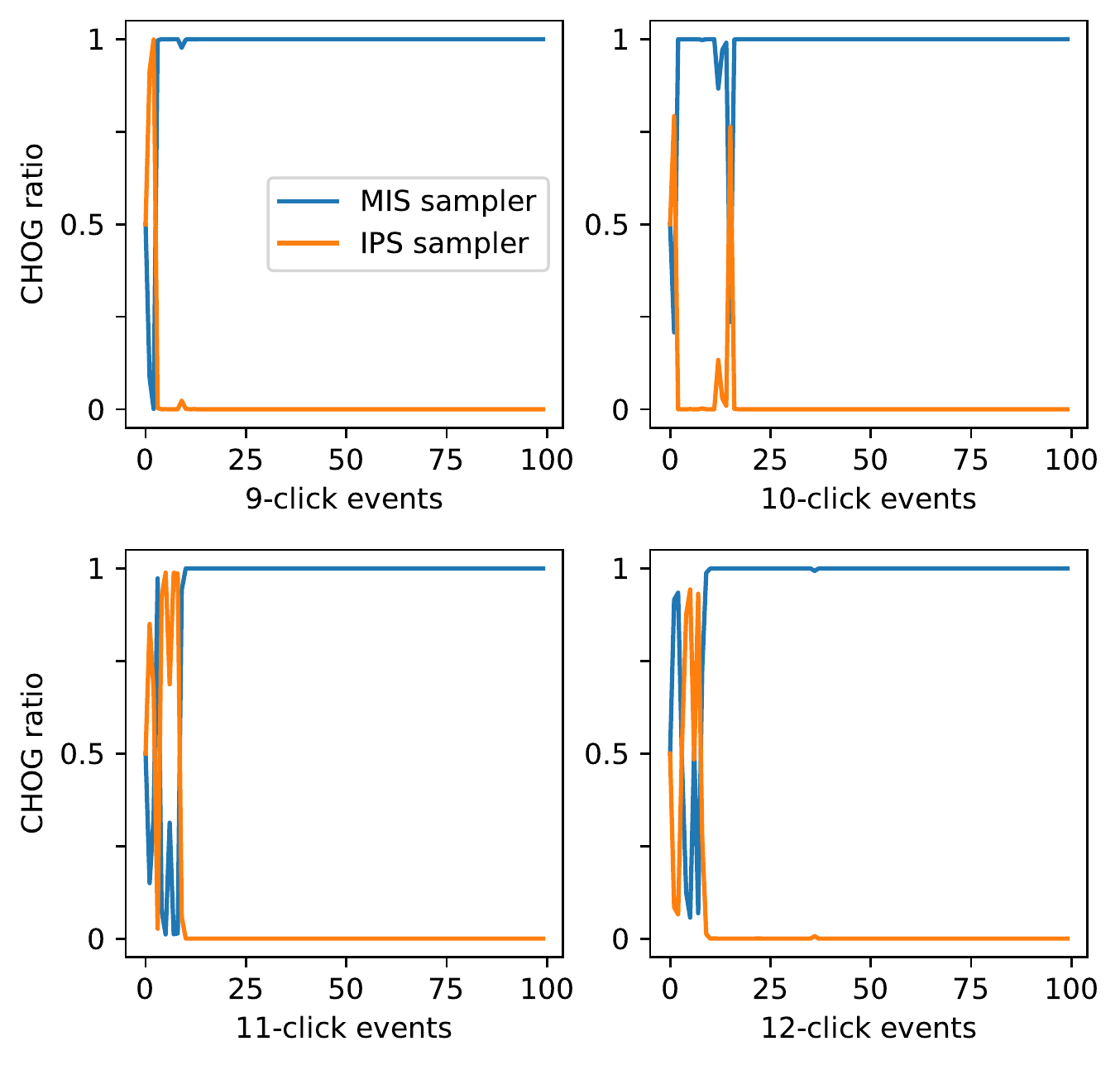}
    \caption{$M=24$ CHOG tests for different total click numbers to determine the relative likelihood of MIS and IPS samples from the ideal click distribution. Convergence to 1 for the MIS samplers indicates they are more likely to have come from the click distribution than the IPS samples.}
    \label{chog_mis_ips}
\end{figure}

\subsubsection{MIS vs IPS}
Our MIS method takes the IPS as its proposal distribution and should then converge to the target (Torontonian) distribution. Here, we use a CHOG ratio test to validate trial MIS samples against the IPS distribution as the adversary. Our discussion in the previous section showed that IPS samples are more likely to be drawn from the ideal distribution than thermal samples, and so here they should provide a more stringent test.

We use a covariance matrix corresponding to 6 sources of two-mode squeezed vacuum, with squeezing parameter $r=1.55$ and transmission $\eta=0.3$, injected into a Haar random 24-mode unitary interferometer. We draw $10^5$ samples from the IPS distribution and use these in an MIS chain with burn-in of 50 and a thinning interval of 10. We then post-select for samples of different fixed total click numbers and update the CHOG ratio. Results for click numbers of between 9 and 12 are shown in Fig.~\ref{chog_mis_ips}. The convergence to 1 for the MIS samples shows that they are more likely to have been drawn from the ideal distribution than the starting IPS samples, and this is indicative of convergence of the chain.

\subsection{Two-point correlators}
Two-point correlators have been proposed as a benchmark for GBS~\cite{dphil_tpc}. Two-point correlations of the light emerging from some optical network are defined as:

\begin{equation}
    C_{i,j}=\langle \Pi_{1}^{i}\Pi_{1}^{j}\rangle - \langle \Pi_{1}^{i}\rangle\langle\Pi_{1}^{j}\rangle,
\end{equation}
where the projector $\Pi_{1}^{i}=\mathbb{I}-\ket{0}_{i}\bra{0}_{i}$ corresponds to a click on mode $i$.

The distributions of two-point correlators are expected to differ between ideal squeezed and thermal samplers. Correlators from a GBS device can therefore be used to validate a squeezed over a thermal hypothesis. As discussed in the main text, the IPS distribution naturally includes interference of pairs of photons. We therefore expect the two-point correlators for this distribution to match those for the ideal distribution.

\begin{figure}[t]
    \centering
    \includegraphics[width=0.45\textwidth]{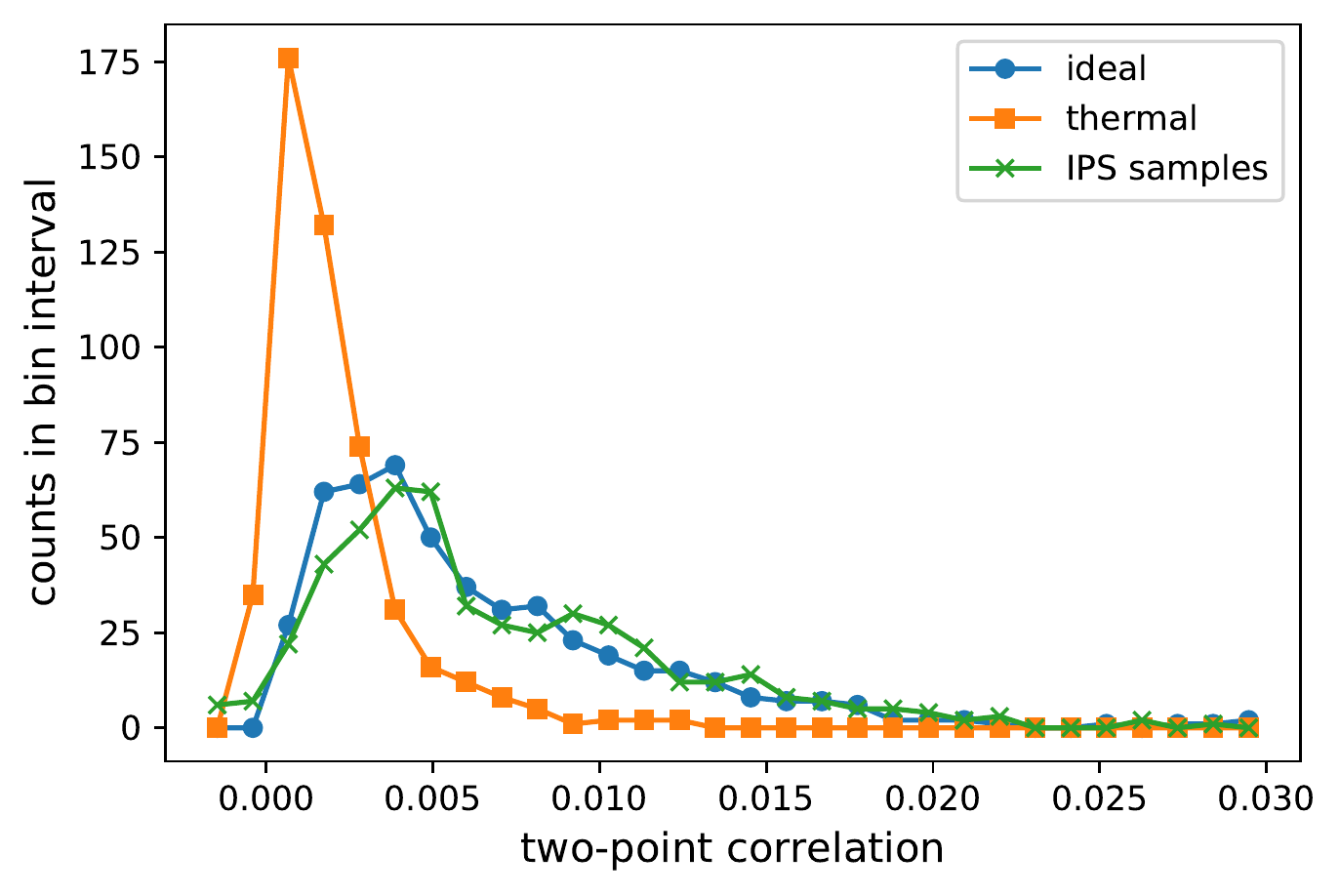}
    \caption{$M=32$ histogram of two-point correlator values for ideal (squeezed), thermal and IPS distributions. The values for ideal and IPS show good overlap and differ significantly from those for the thermal distribution.}
    \label{tpc_ideal_ips_thermal}
\end{figure}

For the ideal and IPS distributions we use a covariance matrix corresponding to 8 sources of two-mode squeezed vacuum, with squeezing parameter $r=1.55$ and transmission $\eta=0.3$, injected into a Haar random 32-mode unitary interferometer. We draw $10^5$ IPS samples for each single output and pair of output modes, convert them to click patterns and use these to estimate the IPS click probabilities and evaluate $C_{i,j}$. For the thermal distribution we use the same unitary interferometer and transmissions but set the mean photon number for 16 thermal states to be $\sinh^{2}(r)$. Results are shown in Fig.~\ref{tpc_ideal_ips_thermal}.

The two-point correlators for the squeezed and IPS distributions are in good agreement. Slight deviations arise from probability estimation errors due to finite sampling. These distributions both significantly diverge from that for the thermal correlators. The IPS distribution is efficient to sample and shows high overlap with the ideal distribution for squeezers, suggesting such tests are not a sufficient indicator of GBS complexity.

\bibliographystyle{apsrev4-1}
\bibliography{bib.bib}

\end{document}